\documentclass[twocolumn, 8pt]{article}
\usepackage{amsfonts,amssymb,latexsym,amsmath,amscd}
\usepackage{times,mathptm}
\usepackage[ps2pdf,colorlinks]{hyperref}
\usepackage{anysize}
\marginsize{.7in}{.3in}{0.7in}{1.3in}

\newcommand{\ket}[1]{|#1\rangle}
\newcommand{\bra}[1]{\langle#1|}

\title{Asymptotically Optimal Circuits \\ for Arbitrary $n$-qubit
       Diagonal Computations\thanks{
        Partially supported
        by the University of Michigan mathematics department VIGRE
        grant and the DARPA QuIST program.
        The views and conclusions contained herein are those of the authors
        and should not be interpreted as necessarily representing official
        policies or endorsements, either expressed or implied,
        of employers and funding agencies.
        }
       }
\author{Stephen S. Bullock and Igor L. Markov\\
        N.I.S.T.-Gaithersburg \& The University of Michigan \\
        {\tt stephen.bullock@nist.gov \quad imarkov@eecs.umich.edu}}

\newenvironment{example}
    {
    \smallskip
    \refstepcounter{theorem}
    \noindent
    {\bf Example \arabic{section}.\arabic{theorem}} \ \ }
    {\hspace*{\fill}{$\Diamond$}
    \smallskip}

\newenvironment{remark}
    {
    \smallskip
    \refstepcounter{theorem}
    \noindent
    {\bf Remark \arabic{section}.\arabic{theorem}} \ \ }
    {\hspace*{\fill}{$\Diamond$}
    \smallskip}

\newenvironment{definition}
    {
    \smallskip
    \refstepcounter{theorem}
    \noindent
    {\bf Definition \arabic{section}.\arabic{theorem}} \ \ }
    {\hspace*{\fill}{\ }
    \smallskip}

\newenvironment{proof}[1][]
    {
    \noindent
    {\bf Proof{#1}:  }
    }
    {\hspace*{\fill}{$\Box$}\smallskip}

\newenvironment{sketch}[1][]
    {
    \noindent
    {\bf Sketch{#1}:  }
    }
    {\hspace*{\fill}{$\Box$}\smallskip}

    {
    \noindent
    {\bf Reference:  }
    }
    {\hspace*{\fill}{$\odot$}\smallskip}

\newtheorem{theorem}{Theorem}[section]
\newtheorem{proposition}[theorem]{Proposition}
\newtheorem{lemma}[theorem]{Lemma}
\newtheorem{corollary}[theorem]{Corollary}

\begin{document}

\newsavebox{\Wgate}
\savebox{\Wgate}(0,0){
    \begin{picture}(0,0)
            \put(0.5,0.5){\framebox(1,1){\large W}}
            \put(0,1){\line(1,0){0.5}}
            \put(1.5,1){\line(1,0){0.5}}
    \end{picture}
}

\newsavebox{\Xgate}
\savebox{\Xgate}(0,0){
    \begin{picture}(0,0)
            \put(0.5,0.5){\framebox(1,1){\large X}}
            \put(0,1){\line(1,0){0.5}}
            \put(1.5,1){\line(1,0){0.5}}
    \end{picture}
}

\newsavebox{\Vgate}
\savebox{\Vgate}(0,0){
    \begin{picture}(0,0)
            \put(0.5,0.5){\framebox(1,1){V}}
            \put(0,1){\line(1,0){0.5}}
            \put(1.5,1){\line(1,0){0.5}}
    \end{picture}
}

\newsavebox{\Rzgate}
\savebox{\Rzgate}(0,0){
    \begin{picture}(0,0)
            \put(0.2,0.2){\framebox(1.6,1.6){{$R_z        $}}}
            \put(0,1){\line(1,0){0.2}}
            \put(1.8,1){\line(1,0){0.2}}
    \end{picture}
}

\newsavebox{\hwire}
\savebox{\hwire}(0,0)
{
    \begin{picture}(0,0)
    \put(0,1){\line(1,0){2}}
    \end{picture}
}

\newsavebox{\botCNOT} 
\savebox{\botCNOT}(0,0){
    \begin{picture}(0,0)
            \put(1,3){\circle{1}}
            \put(0,3){\line(1,0){2}}
            \put(1,1){\line(0,1){2.5}}
        \put(1,1){\circle*{.4}}
    \end{picture}
}

\newsavebox{\topCNOT} 
\savebox{\topCNOT}(0,0){
    \begin{picture}(0,0)
            \put(1,1){\circle{1}}
            \put(0,3){\line(1,0){2}}
            \put(1,.5){\line(0,1){2.5}}
            \put(1,3){\circle*{.4}}
        \put(0,1){\line(1,0){2}}
    \end{picture}
}

\newsavebox{\longCNOT} 
\savebox{\longCNOT}(0,0){
    \begin{picture}(0,0)
            \put(1,1){\circle{1}}
            \put(0,3){\line(1,0){2}}
        \put(0,5){\line(1,0){2}}
            \put(1,.5){\line(0,1){4.5}}
            \put(1,5){\circle*{.4}}
            \put(0,1){\line(1,0){2}}
    \end{picture}
}

\maketitle

\begin{abstract}
A unitary operator $U=\sum_{j,k} u_{j,k} \ket{k} \bra{j}$ is called
\emph{diagonal} when $u_{j,k}=0$ unless $j=k$.  The definition extends
to quantum computations, where $j$ and $k$ vary over the $2^n$ binary
expressions for integers $0,1 \cdots ,2^n-1$, given $n$ qubits.  Such
operators do not affect outcomes of the projective measurement
$\{ \bra{j} \; ; \; 0 \leq j \leq 2^n-1\}$ but rather
create arbitrary relative phases among the computational basis states
$\{ \ket{j} \; ; \; 0 \leq j \leq 2^n-1\}$.  These relative phases
are often required in applications.

Constructing quantum circuits
for diagonal computations using standard techniques
requires either $O(n^2 2^n)$
controlled-not gates and one-qubit Bloch sphere rotations or else
$O (n 2^n)$ such gates and a work qubit.  This work provides a
recursive, constructive procedure which inputs the matrix coefficients
of $U$ and outputs such a diagram containing $2^{n+1}-3$ alternating
controlled-not gates and one-qubit $z$-axis Bloch sphere rotations.
Up to a factor of two, these circuits are the smallest possible.
Moreover, should the computation $U$ be a tensor of diagonal
one-qubit computations of the form
$R_z(\alpha)=\mbox{e}^{-i \alpha/2}\ket{0}\bra{0}+
\mbox{e}^{i \alpha/2} \ket{1} \bra{1}$, then a cancellation of
controlled-not gates reduces our circuit to that of an $n$-qubit tensor.
\end{abstract}


\section{Introduction}
\label{sec:intro}
Let $U(N)=\{ V \mbox{ an } N \times N \mbox{ matrix} \;
; \; V V^* = {\bf 1}\}$, where ${\bf 1}$ is an identity matrix and
$V^* = \bar{V}^t$ is the mathematical notation for the adjoint.
One may view $U(N)$ as the set of all reversible quantum computations
acting on $n$ qubits.
Then our usual convention is that algorithms for \emph{quantum circuit
synthesis} input such a $V \in U(N)$ and output
a quantum circuit diagram for $V$, up to global phase.
Several distinct quantum circuits may realize the same
computation $V$.  Thus, one seeks circuits for which the total number of
gates is small.  This work
focuses on the case where the input computation is
diagonal.

Gate counts for quantum circuits are often made in terms of
\emph{basic gates} \cite{BarencoEtAl:95}, i.e., the set of all controlled-not
gates and one-qubit computations.  Our gate counts will be made with respect
to the following gate library.  We refer to 
elements as \emph{elementary gates}, in contrast to basic gates.
\begin{enumerate}
\item
\label{gate:Ry}
For $1 \leq j \leq n$, apply $R_y (\theta) \in U(2^1)$ on line $j$,
where
\begin{equation}
\begin{array}{lcl}
R_y(\theta) & =
& \cos \frac{\theta}{2} \; \ket{0} \bra{0} \; +  
\sin \frac{\theta}{2} \; \ket{0} \bra{1} \\
& & - \sin \frac{\theta}{2} \; \ket{1} \bra{0} +  
\cos \frac{\theta}{2} \; \ket{1} \bra{1} ,
\; 0 \leq \theta < 2 \pi \\
\end{array}
\end{equation}
is a $y$-axis Bloch sphere rotation
\cite[\S4.2]{NielsenC:00}.

\item
\label{gate:Rz}
For $1 \leq j \leq n$, apply $R_z (\alpha) \in U(2^1)$ on line $j$,
where
\begin{equation}
R_z(\alpha)=
\mbox{e}^{-i \alpha/2} \; \ket{0} \bra{0}
+ \mbox{e}^{i \alpha/2} \ket{1} \bra{1}, \quad 0 \leq \theta < 2 \pi
\end{equation}
is a $z$-axis Bloch sphere rotation
\cite[\S4.2]{NielsenC:00}.

\item
\label{gate:CNOT}
Let $1 \leq j,k \leq n$, let $b_1, b_2, \cdots, b_n$ be $n$ variables
varying in the field of two elements $\mathbb{F}_2$, and let
$x,y \mapsto x \oplus y$ denote the exclusive-or ({\tt XOR}) operator
which is addition in $\mathbb{F}_2$.  The final type of elementary gate
is the $j$-controlled-not gate acting on line $k$.  We denote it
by $C_j^k$.  In case $j<k$,
\end{enumerate}
\begin{equation}
C_j^k = \sum_{0 \leq b_1\cdots b_n \leq N-1}
\ket{b_1 \cdots b_{j} \cdots (b_j \oplus b_k) \cdots b_n}
\bra{b_1 \cdots b_{j} \cdots b_k \cdots b_n}
\end{equation}

\noindent
\quad \quad The other case $k<j$ is similar.

\noindent
The elementary gate library is universal because
any $V \in U(N)$ factors into basic gates \cite{BarencoEtAl:95} and any
one-qubit computation $W$ can be decomposed
into $W = \mbox{e}^{i \Phi} R_y(\theta_1)
R_z(\alpha) R_y(\theta_2)$ for $\mbox{e}^{i\Phi}$ an unmeasurable global
phase \cite[Lemma4.1]{BarencoEtAl:95} \cite[\S4.2]{NielsenC:00}.
Moreover, the
asymptotics $\Omega(-)$,
$O(-)$, and $\Theta(-)$ of the counts in either gate library are
identical, since every elementary gate is basic while every basic gate
factors into at most three elementary gates.

We next set some conventions.
Throughout, $U$ is a diagonal quantum computation on $n$ qubits.
Thus, for $N=2^n$, $U$ acts on the $n$-qubit state space which is the
$\mathbb{C}$ span of the computational basis $\{ \ket{j} \; ; \; 0 \leq
j \leq N-1 \}$.  The $j$ are typically written as binary integers.
As $U$ is diagonal, $U=\sum_{j=0}^{N-1} u_j \ket{j} \bra{j}$.  Moreover,
$U$ unitary implies $|u_j|^2=1$.

We denote the Lie group \cite{Knapp:98} of all diagonal computations
on $n$-qubit states by $A$.  The notation $A(n)$ may be used for
emphasis.  Observe that $A$ is abelian, i.e., commutative.

Circuit synthesis algorithms that provably produce minimal gate counts are rare,
difficult to construct, and have been published for special cases
only \cite{SongK:03}. Before stating our main result,
we formalize a sense in which it is best possible.

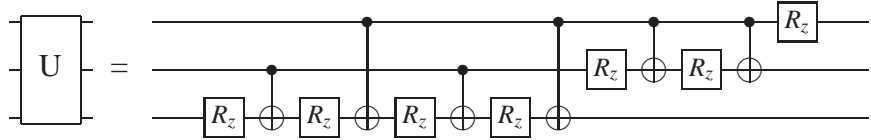
\begin{figure*}[t]

\begin{center}
\begin{picture}(36,6)

\put(0,1){\line(1,0){.5}}
\put(0,3){\line(1,0){.5}}
\put(0,5){\line(1,0){.5}}
\put(.5,.75){\line(0,1){4.5}}
\put(3,.75){\line(0,1){4.5}}
\put(.5,5.25){\line(1,0){2.5}}
\put(0.5,0.75){\line(1,0){2.5}}
\put(3,1){\line(1,0){.5}}
\put(3,3){\line(1,0){.5}}
\put(3,5){\line(1,0){.5}}
\put(1.25,2.75){\large U}

\put(4.25,2.75){\large =}

\put(6,0){\usebox{\hwire}}
\put(6,2){\usebox{\hwire}}
\put(6,4){\usebox{\hwire}}

\put(8,0){\usebox{\Rzgate}}
\put(8,2){\usebox{\hwire}}
\put(8,4){\usebox{\hwire}}

\put(10,0){\usebox{\topCNOT}}
\put(10,4){\usebox{\hwire}}

\put(12,0){\usebox{\Rzgate}}
\put(12,2){\usebox{\hwire}}
\put(12,4){\usebox{\hwire}}

\put(14,0){\usebox{\longCNOT}}

\put(16,0){\usebox{\Rzgate}}
\put(16,2){\usebox{\hwire}}
\put(16,4){\usebox{\hwire}}

\put(18,0){\usebox{\topCNOT}}
\put(18,4){\usebox{\hwire}}

\put(20,0){\usebox{\Rzgate}}
\put(20,2){\usebox{\hwire}}
\put(20,4){\usebox{\hwire}}

\put(22,0){\usebox{\longCNOT}}

\put(24,0){\usebox{\hwire}}
\put(24,2){\usebox{\Rzgate}}
\put(24,4){\usebox{\hwire}}

\put(26,0){\usebox{\hwire}}
\put(26,2){\usebox{\topCNOT}}

\put(28,0){\usebox{\hwire}}
\put(28,2){\usebox{\Rzgate}}
\put(28,4){\usebox{\hwire}}

\put(30,0){\usebox{\hwire}}
\put(30,2){\usebox{\topCNOT}}

\put(32,0){\usebox{\hwire}}
\put(32,2){\usebox{\hwire}}
\put(32,4){\usebox{\Rzgate}}

\put(34,0){\usebox{\hwire}}
\put(34,2){\usebox{\hwire}}
\put(34,4){\usebox{\hwire}}

\end{picture}
\end{center}

\caption{ \label{fig:3qcase}
This diagram shows the circuit structure realizing a three
qubit diagonal computation using our circuit synthesis algorithm.
The general algorithm applies in $n$ rather than merely three qubits
and extends the construction of Section 2.2 of a previous work
\cite{BullockMarkov:02}.  Should the input diagonal be of the form
$U=R_z(\alpha_1) \otimes R_z(\alpha_2) \otimes R_z(\alpha_3)$,
the second, third, fourth, and sixth $R_z$ gates of the output diagram
are trivial, implying that all controlled-not gates cancel.  The output
then coincides with the input.}
\end{figure*}

\begin{definition}
Let $H \subset U(N)$ be an analytic subgroup.  An $n$-qubit
quantum circuit synthesis algorithm with inputs restricted to $H$
is said to stably output to $H$ iff i) it outputs at most a countably
infinite number of quantum circuit topologies containing only
elementary gates
as inputs are varied over all of $H$ and ii) for each such
circuit topology $\tau$, the corresponding computations remain in
$H$ for every variation of parameter on any $R_y(\theta)$, $R_z(\alpha)$
gate within $\tau$.  If $\varsigma$ is such a synthesis algorithm
accepting any input from $H$ and outputting
stably to $H$, we put $\# \varsigma = \mbox{max} \{ \# \tau \; ; \;
\tau \mbox{ is a diagram output by } \varsigma\}$, where $\# \tau$ refers
to the number of elementary gates in $\tau$.
We finally put
\begin{equation}
\ell (H) = \mbox{min} \{ \# \varsigma \; ; \; \varsigma \mbox{ outputs
stably to } H \}
\end{equation}
\end{definition}

\begin{definition}
\label{def:stablyasymptoticallyoptimal}
Consider now a family $\{H(n) \subset U(2^n) \; ; \; n \geq 1\}$ of
analytic subgroups \cite[p. 47]{Knapp:98}.
A family of $n$-qubit synthesis algorithms $\{\varsigma(n) \; ; \; n \geq 1\}$,
each allowing for any input in $H(n)$ and outputting stably to $H(n)$,
will be called \emph{stably asymptotically optimal} iff
$\# \varsigma(n) \in O(\ell[H(n)])$.
\end{definition}

\begin{theorem}
\label{thm}
Any $n$-qubit diagonal computation $U \in A(n)$ may be realized
by a quantum circuit holding $2^{n+1}-3$ alternating
controlled-not gates and $z$-axis Bloch sphere rotations $R_z(\theta)$.
The construction is stably asymptotically optimal for $A(n)$.
\end{theorem}

\begin{remark}
Two other comments should be made about the construction.  First, it
requires neither a work qubit \cite{BarencoEtAl:95}
nor any $R_y(\theta)$ elementary gates.
Second, should a $n$-qubit tensor of the form $\otimes_{j=1}^n R_z(\alpha_j)$
be input to the algorithm, the output will hold several cancelling
controlled-not gates.  After cancellation, the output
will match the input.
\end{remark}

As a benchmark, we describe in Section \ref{sec:prior}
a diagram for a given diagonal $U$ using standard techniques.
The technique hinges on a
well-known circuit diagram for an $(n-1)$-qubit controlled
element of $A(1)$.
In the presence of one ancilla (work) qubit, this diagram holds
$O(n 2^n)$ basic or equivalently
elementary gates.  The cost rises to $O (n^2 2^n)$ when there is
no ancilla qubit.  Thus, the asymptotic cost of $O (2^n)$
of the synthesis algorithm of the Theorem (see Section
\ref{sec:xr}) compares
favorably with known results.  Moreover, dimension counts
during the argument for
stably asymptotically optimal will make clear that synthesizing
large subsets of $A$ requires $\geq 2^n -1$ elementary
gates. In this specialized sense, $\Omega(2^n)$ gates are required, and
the diagram of Section \ref{sec:xr} proves that
diagrams for generic diagonal computations cost $\Theta(2^n)$ elementary
(or basic) gates.

See Figure \ref{fig:3qcase} for the overall circuit topology in the case
$n=3$ qubits.  We defer a description of the algorithm for computing
the $R_z$ angles to the body
and next discuss potential applications.

The first application is in conjunction with the standard
synthesis algorithm \cite{BarencoEtAl:95,Cybenko:01}
\cite[\S4.5]{NielsenC:00},
which may be formalized using the $QR$ matrix decomposition
\cite{GolubVL:96,Cybenko:01}.  For $V \in U(N)$, the
algorithm uses a matrix
factorization $V=QR$, where $Q$ is a product of Givens
rotations \cite{Cybenko:01} realizable as $(n-1)$-controlled one-qubit
computations and $R$ is diagonal. Should the projective measurement
$\{ \bra{j} \; ; \; 0 \leq j \leq N-1\}$ follow $V$, one need not apply $R$.

Consider instead the following situation.  For $p<<n$, a desired computation
$V\in U(N)$ is known to arise from $V_1, V_2, \cdots, V_{n-p+1} \in U(2^p)$
as follows.  First $V_1$ is applied on lines $1,2, \cdots ,p$,
after which $V_2$ is applied on lines $2,3,\cdots, p+1$, and so on until
finally $V_{n-p+1}$ is applied on lines $n-p+1, n-p+2, \cdots n$.  If
quantum computing technology has progressed so that $O(np2^p)$
elementary gates may be realized directly, one may factor
each $V_1 = Q_1 R_1$, $V_2=Q_2 R_2$, \dots $V_{n-p+1}=Q_{n-p+1}R_{n-p+1}$
and apply the standard synthesis algorithm on each subblock.  However,
with the convention that $\{ \bra{j} \; ; \; 0 \leq j \leq N-1 \}$ is only
applied after the entire computation $V$, we now need quantum circuits
realizing each of the $R_1, R_2, \cdots, R_{n-p+1}$.  The synthesis
algorithms proposed in this paper
provide these.  Moreover, note that essential part of the argument
is merely the overlap of the smaller blocks, not their pattern.

Two further instances commonly arise where one needs to be careful
about relative phases of computational basis states.
\begin{itemize}
\item  Suppose that for $V \in U(2^{n-1})$, one wishes to build a
circuit for the computation $({\bf 1} \oplus V) \in U(2^n)$ which
applies $V$ iff the top line carries $\ket{1}$.  Suppose one has
a circuit for $V$, correct up to relative phase.  For example, such
results from the factorization of $Q$ into Givens rotations using
$V=QR$ \cite{Cybenko:01}.   A straightforward approach is to
condition every gate in $Q$, so that e.g. conditioned-not gates in
$Q$ correspond to Toffoli computations in ${\bf 1} \oplus Q$.  Yet
${\bf 1} \oplus R$ will affect measurements in the $n$-qubit
computational basis, unlike the diagonal computation $R$ in
$(n-1)$ qubits.  One even needs a conditioned gate for the \emph{global}
phase of the original $V$.
\item  Moreover, circuits for diagonal computations are required
whenever the final projective measurement \cite[\S2.2.5]{NielsenC:00}
is not $\{ \bra{j} \; ; \; 0 \leq j \leq N-1\}$.
\end{itemize}

Another possible application of circuits for diagonal quantum computations
is to reduce the synthesis of arbitrary quantum
computations to the synthesis of real quantum computations \cite{YShi:03},
i.e., of those
$V \in O(N)=\{ V \in U(N) \; ; \; V = \bar{V} \}$.  For there is a matrix
decomposition $U(N) = O(N) \; A \; O(N)$.  Indeed, this is
a special case of the
$KAK$ metadecomposition \cite{Knapp:98, KhanejaBG:01a, BullockMarkov:02}.
Thus if
$V \in U(N)$ is arbitrary, we may write $V=O_1 U O_2$ for
$O_1, O_2 \in O(N)$ real quantum computations and $U \in A(n)$.  The
present work produces a circuit for $U \in A(n)$, reducing the question
of a circuit for $V \in U(N)$ to circuits for $O_1, O_2 \in O(N)$.

Finally, we expect further applications to other quantum circuit synthesis
algorithms relying on other examples of the $KAK$ matrix metadecomposition.
Another such example is the {\tt Cosine-Sine} decomposition\cite{Tucci:99}.
This decomposition states that one may write any $V \in U(N)$ as
$V= (U_1 \oplus U_2) W (U_3 \oplus U_4)$ for
$U_1,U_2,U_3,U_4 \in U(N/2)$ and $W$ a sparse matix whose
nonzero entries are paired cosines and sines.  A quantum circuit
for the matrix $W$ may be
synthesized using the algorithm of this paper.  Indeed, let
$S= \ket{0} \bra{0} + i \ket{1} \bra{1}$ and
$H$ denote the Hadamard gate, costing one and two elementary
gates respectively \cite{BullockMarkov:02}.  Then for ${\bf 1}$ an
$(N/2) \times (N/2)$ identity matrix, one may compute that
$U= [SH \otimes {\bf 1}] W [(SH)^* \otimes {\bf 1}] \in A$ is a
diagonal computation.  Hence, one
may implement the nonrecursive portion of {\tt Cosine-Sine} synthesis
using the methods of this paper and six extra elementary gates.

We briefly outline the body of the paper.  Section \ref{sec:prior}
describes an algorithm for building quantum circuits for diagonal
computations which is analogous to an unoptimized version of classical
two-level synthesis of logic functions.  This algorithm
produces $O(n2^n)$ gates with a single ancilla qubit
and $O(n^2 2^n)$ gates else.  Section \ref{sec:chi} outlines how to
use Lie theory \cite{Knapp:98} to recognize when $U_n \in A(n)$ factors
as a tensor on line $n$, i.e., case
$U_n = U_{n-1} \otimes R_z(\alpha)$ for $U_{n-1} \in A(n-1)$.
Section \ref{sec:xr} motivates and describes the
recursive construction of the circuits of Theorem \ref{thm}.
Finally, Section \ref{sec:lower}
discusses dimension counts required for the lower bounds proving that
our circuit diagrams are generically asymptotically optimal.
Appendix \ref{sec:cr} gives a construction similar to that of the
Theorem, using $(n-1)$-controlled $R_z$ gates.

Finally, some mathematical background beyond that usually associated
to the quantum computing literature \cite{NielsenC:00} is required
to understand the arguments in this manuscript.  The constructive synthesis
algorithm makes use of the Lie theory of commutative
matrix groups \cite{Knapp:98}.  The argument for stable lower bounds makes
use of the theory of smooth manifolds as is commonly treated in
differential topology \cite{GuilleminPollack:74}.

\section{Prior Work}
\label{sec:prior}

\subsection*{Circuits with measurement gates of Hogg et al.}

Hogg et al. \cite{HoggMPR:98} consider synthesis of quantum
circuits for diagonal computations from a much different
perspective. Their main result is polynomial-size circuits, but in
somewhat different circumstances compared to our work.

\begin{itemize}
\item  The diagonal computations
$U=\sum_{j=0}^{N-1} u_j \ket{j} \bra{j}$ to which the prior result applies
are required to have many $u_j$ repeat.  Indeed, accounting for the global
phase, one supposes a family of diagonal computations
$\{ U_n = \sum_{j=0}^{N-1} u_{n,j} \ket{j} \bra{j} \; ; \; n \geq 1 \}$
where $\# \{u_{n,j} \neq 1 \; ; \; 0 \leq j \leq 2^n -1 \}$
scales as some polynomial $p(n)$.
\item  Moreover, the algorithm chosen in later steps depends on outputs
of measurements of the quantum memory state in earlier steps.  In the
construction of classical circuits, the gate
count would be increased by at least one {\tt MUX} (if-then-else) gate
for each classical branching, and each unique $u_j$ contributes
such a branching.  The presense of measurement gates moreover takes
their algorithm out of the present context of reversible gate libraries.
\item  The circuits ibid. would be large on a generic input of
$\otimes_{j=1}^n R_z(\alpha_j)$ due to little repetition in the input
phases.  Thus, a separate section \cite[\S4]{HoggMPR:98} describes a
precomputation to determine whether an input is of the form
$\otimes_{j=1}^n R_z(\alpha_j)$.  If
this is the case, one should instead choose the tensor diagram.  In
contrast, given an input $U=\otimes_{j=1}^n R_z(\alpha_j)$, our output
circuits automatically contain several cancelling controlled-nots's.
After cancellation, one recovers the input tensor.
\end{itemize}

Despite these caveats, the citation above does include some of the
discussion of the next subsection.

\ \\
\ \\
\ \\

\subsection*{Analogies to classical two-level logic}

\begin{figure}
\begin{center}
\begin{picture}(21,10.5)
\put(1,0){\usebox{\hwire}}
\put(1,2){\usebox{\hwire}}
\put(1,4){\usebox{\Xgate}}
\put(1,6){\usebox{\Xgate}}
\put(1,8){\usebox{\hwire}}
\put(0,9){$\ket{0}$}

\put(3,0){\usebox{\Vgate}}
\put(3,2){\usebox{\hwire}}
\put(3,4){\usebox{\hwire}}
\put(3,6){\usebox{\hwire}}
\put(3,8){\usebox{\hwire}}

\put(4,3){\circle*{0.4}}
\put(4,5){\circle*{0.4}}
\put(4,7){\circle*{0.4}}
\put(4,1.5){\line(0,1){5.5}}

\put(5,0){\usebox{\hwire}}
\put(5,2){\usebox{\hwire}}
\put(5,4){\usebox{\Xgate}}
\put(5,6){\usebox{\Xgate}}
\put(5,8){\usebox{\hwire}}

\put(8,4){\LARGE $=$}

\put(11,0){\usebox{\hwire}}
\put(11,2){\usebox{\hwire}}
\put(11,4){\usebox{\Xgate}}
\put(11,6){\usebox{\Xgate}}
\put(11,8){\usebox{\hwire}}
\put(10,9){$\ket{0}$}

\put(14,9){\circle{1}}
\put(14,7){\circle*{0.4}}
\put(14,5){\circle*{0.4}}
\put(14,3){\circle*{0.4}}
\put(14,3){\line(0,1){6.5}}

\put(13,0){\usebox{\hwire}}
\put(13,2){\usebox{\hwire}}
\put(13,4){\usebox{\hwire}}
\put(13,6){\usebox{\hwire}}
\put(13,8){\usebox{\hwire}}

\put(16,9){\circle*{0.4}}
\put(16,1.5){\line(0,1){7.5}}

\put(15,0){\usebox{\Vgate}}
\put(15,2){\usebox{\hwire}}
\put(15,4){\usebox{\hwire}}
\put(15,6){\usebox{\hwire}}
\put(15,8){\usebox{\hwire}}

\put(18,9){\circle{1}}
\put(18,7){\circle*{0.4}}
\put(18,5){\circle*{0.4}}
\put(18,3){\circle*{0.4}}
\put(18,3){\line(0,1){6.5}}

\put(17,0){\usebox{\hwire}}
\put(17,2){\usebox{\hwire}}
\put(17,4){\usebox{\hwire}}
\put(17,6){\usebox{\hwire}}
\put(17,8){\usebox{\hwire}}

\put(19,0){\usebox{\hwire}}
\put(19,2){\usebox{\hwire}}
\put(19,4){\usebox{\Xgate}}
\put(19,6){\usebox{\Xgate}}
\put(19,8){\usebox{\hwire}}
\end{picture}
\end{center}

\caption{
\label{fig:ErsatzAND}
For $S=\{1,2\} \subset \{1,2,3\}$, this figure shows at left
$[X^{\delta_S} \otimes {\bf 1}]
\circ \Lambda_{\{1,2,3\}}(V) \circ [X^{\delta_S} \otimes {\bf 1}]$.  At right
is the first reduction of this circuit in a common implementation
\cite{BarencoEtAl:95}.
}
\end{figure}
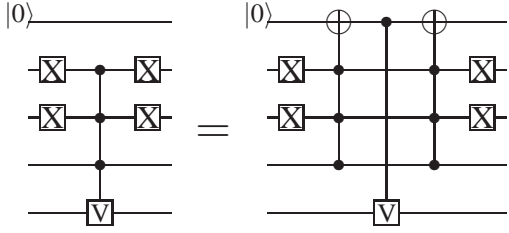

We briefly recall classical two-level synthesis in order to contrast
our circuits with this technique.  Thus,
let $\mathbb{F}_2$ denote the field of two elements, and $b \in \mathbb{F}_2$
also denote either a Boolean value or an integer of $\{0,1\}$.
In this section, let $\bar{b} = b_1 b_2 \cdots b_n \in (\mathbb{F}_2)^n$
denote an $n$-bit string.  Suppose $\varphi: (\mathbb{F}_2)^n \rightarrow
\mathbb{F}_2$ describes an $n$-to-$1$ Boolean function we wish to
realize with a circuit in the classical, irreversible
{\tt AND}-{\tt OR}-{\tt NOT} gate library.  A textbook technique
\cite{HachtelS:00} is the two-level approach.  Briefly, take
$b_1, b_2, \cdots b_n$ as variables, and let
$\bar{c}=c_1c_2 \cdots c_n \in (\mathbb{F}_2)^n$ be a fixed bit string
with $\varphi(\bar{c})=1$.  Denote by $\delta_{\bar{c}}$ the indicator
function of $\bar{c}$, i.e., $\delta_{\bar{c}}:(\mathbb{F}_2)^n \rightarrow
\mathbb{F}_2$ has $\delta_{\bar{c}}(\bar{c})=1$ and
$\delta_{\bar{c}}(\bar{b})=0$ for $\bar{b} \neq \bar{c}$.
Then we have
\begin{equation}
\delta_{\bar{c}}=
[{\tt NOT}^{c_1\oplus 1}(b_1)] {\tt AND}
[{\tt NOT}^{c_2\oplus 1}(b_2)] {\tt AND} \cdots
[{\tt NOT}^{c_n\oplus 1}(b_n)],
\end{equation}
where the ${\tt AND}$ gates are equivalently multiplication
in $\mathbb{F}_2$.  Moreover, if
$\{ \bar{c}_1, \bar{c}_2, \cdots , \bar{c}_\ell \} =
\{ \bar{b} \in (\mathbb{F}_2)^n \; ; \; \varphi(\bar{b})=1 \}$, then
the expression
\begin{equation}
\varphi = \; \delta_{\bar{c}_1} \; {\tt OR} \; \delta_{\bar{c}_2}
\; {\tt OR} \cdots
{\tt OR} \; \delta_{\bar{c}_\ell}
\end{equation}
provides an ${\tt AND}-{\tt OR}-{\tt NOT}$ circuit.  For generic $\varphi$
with $\ell \approx 2^{n-1}$, note this classical two-level circuit
requires $O(2^{n-1})$ gates.

Optimizing such two-level circuits is {\tt NP}-hard \cite{HachtelS:00},
and the problem has been studied extensively since the late 1960s.
Algorithms and tools for this problem,
e.g. {\tt Espresso}, are widely known, and some
are used in commercial circuit design tools.  More recently,
two-level decompositions in the {\tt AND}-{\tt XOR}-{\tt NOT}
gate library have been
introduced.  This is still universal, as any $b_1, b_2 \in \mathbb{F}_2$
have
$(b_1$ {\tt OR} $b_2) = b_1 \oplus b_2 \oplus (b_1$ {\tt AND} $b_2)$.
Publicly available tools for such ESOP-decomposition include
{\tt EXORCISM-4} \cite{MishchenkoP:01,SteinbachM:01}.
We mention this transition ${\tt OR}\mapsto {\tt XOR}$ as it is loosely
analogous to our change in strategy from Section \ref{sec:cr} to
Section \ref{sec:xr}.  Other work on {\tt ROM}-based quantum computation
\cite{Travaglione:01} has also made use of ${\tt XOR}$ based two-level
synthesis.

We extend these ideas to build a simple circuit for
$U=\sum_{j=0}^{N-1} u_j \ket{j} \bra{j}$ costing
$O(n2^n)$ elementary (or basic) gates.  Recall standard notation
uses $\Lambda_k (V)$ to denote a $k-$controlled $V$ gate for
$V \in U(2^1)$ \cite{BarencoEtAl:95}.  We extend this
notation slightly, in view of
this section and Appendix \ref{sec:cr}.

\begin{definition}
In $n$-qubits, let $S \subset \{1,2,\cdots n-1\}$ and $V \in U(2^1)$.
Then $\Lambda_S (V)$ denotes the particular instance of
$\Lambda_{\# S}(V)$ controlled by lines $\{ j \in S\}$ and acting
on line $n$.
\end{definition}

\begin{definition}
In $n$ qubits, let $S \subset \{1,2,\cdots n-1\}$.  Then
$\delta_S : (\mathbb{F}_2)^n \rightarrow \mathbb{F}_2$
is given by $\delta_S(j)=1$ iff
$[(j \neq n) \mbox{ and } (j \in S)]$.  For
$X=\ket{1}\bra{0}+\ket{0}\bra{1}$ a Pauli-$X$ gate,
we write $X^{\delta_S} = \otimes_{j=1}^n X^{\delta_S(j)}$.
If $0 \leq j \leq N/2-1$,
then $S(j)\subset \{1,2,\cdots n-1\}$ is the subset
\[S(j) = \{ 0 \leq k \leq N/2-1 \; ; \; c_k =1 \mbox{ for }
j = \bar{c} = c_1 c_2 \cdots c_{n-1} \}\]
Finally, for $S \subset \{1,2,\cdots n-1\}$,
the number $k(S)$ is that integer $k$ such that
$S=S(k)$.
\end{definition}

We now detail one construction of a circuit for
$U= \sum_{j=0}^{N-1} u_j \ket{j} \bra{j}$.
Let $0 \leq k \leq N-1$, and $k \equiv 0 \mbox{mod }2$.
Let $V_k = u_k \ket{0} \bra{0} + u_{k+1}\ket{1} \bra{1}$, a
one-qubit computation.  Label
\begin{equation}
U_k = u_k \ket{k} \bra{k} + u_{k+1} \ket{k+1}\bra{k+1}
+ \sum_{j\neq k, j \neq k+1} \ket{j} \bra{j}
\end{equation}
Then we have the following expression.
\begin{equation}
U_k = [X^{\delta_{S(k/2)}} \otimes {\bf 1}] \;
\Lambda_{\{1,2,\cdots,n-1\}} (V_k)
\; [X^{\delta_{S(k/2)}} \otimes {\bf 1}]
\end{equation}
Moreover, all such $U_k$ commute.  Thus for any enumeration
of subsets $S_1, \cdots, S_{N/2} \subset \{1,2,\cdots n-1\}$,
\begin{equation}
\label{eq:2LevAnal}
\begin{array}{lcl}
U & = & [X^{\delta_{S_1}} \otimes {\bf 1}]
\; \Lambda_{\{1,2,\cdots,n-1\}} (V_{k(S_1)})
\; [X^{\delta_{S_1}} \otimes {\bf 1}] \circ \\
& & \ [X^{\delta_{S_2}} \otimes {\bf 1}] \;
\Lambda_{\{1,2,\cdots,n-1\}} (V_{k(S_2)})
\; [X^{\delta_{S_2}} \otimes {\bf 1}] \circ \cdots \circ \\
& & \ [X^{\delta_{S_{N/2}}} \otimes {\bf 1}]
\; \Lambda_{\{1,2,\cdots,n-1\}} (V_{k(S_{N/2})})
\; [X^{\delta_{S_{N/2}}} \otimes {\bf 1}] \\
\end{array}
\end{equation}
This directly produces a quantum circuit built out of subblocks such
as the one illustrated in Figure \ref{fig:ErsatzAND}.

Before passing to the asympotitcs, we note an optimization.  A
\emph{Grey code} \cite[\S4.5.2]{NielsenC:00}
produces a sequence $S_1, S_2, S_3 \cdots S_{N/2}$
with $\# (S_k \cap S_{k+1}) = 1, 1 \leq k \leq N/2-1$.
Sample Grey codes are recalled with $n-1 = 1,2,3$,
where we write $k(S)$ for each subset:
\begin{equation}
\begin{array}{l}
0,1 \\
00, 01, 10, 11 \\
000, 001, 010, 011, 111, 110, 101, 100 \\
\end{array}
\end{equation}
By using a Grey code in the choice of enumeration of the subset for
equation \ref{eq:2LevAnal}, we obtain a massive cancellation of inverters
leaving only $N/2$ such $X$ gates.

Figure \ref{fig:ErsatzAND} recalls the remaining facts justifying the
$O(n2^n)$ gate count for this synthesis algorithm.  Namely, each of the
$N/2$ computations
$\Lambda_{n-1}(V)$ require $O(n^2)$ basic gates absent an ancilla
qubit or $48n-164$ basic gates with an ancilla qubit present
\cite{BarencoEtAl:95}.  Summing produces asymptotics of
$O(n2^n)$ elementary or basic gates with the ancilla present and
$O(n^2 2^n)$ gates without the ancilla present.  In contrast, the
circuits of Theorem \ref{thm} described in Section \ref{sec:xr} require
no ancilla and cost $O(2^n)$ gates.

\section{Tensors and characters}
\label{sec:chi}

The recursive proccess of the two new synthesis algorithms for
diagonal quantum computations in Section \ref{sec:xr} and Appendix
\ref{sec:cr} both rely on well-known ideas from Lie theory \cite{Knapp:98}.
Specifically, it is typical to study Lie groups and most especially
commutative Lie groups using their character functions.
For $G$ a Lie group, a character is a function
$\chi: G \rightarrow \mathbb{C} - \{0\}$ with
$\chi(gh)=\chi(g)\chi(h)$.  The motivating example is
the following group and character.
\[
\begin{array}{l}
G=GL(n,\mathbb{C}) = \{ M \ n \times n \mbox{ complex matrix}
\; ; \; \exists \; M^{-1} \} \\
\chi = \mbox{det} : GL(n,\mathbb{C}) \rightarrow \mathbb{C}-\{0\} \\
\end{array}
\]
Note that for any character, $\log \chi (gh) = \log \chi (g) \; + \;
\log \chi (h)$ and by continuity $\log \chi (g^a) = a \log \chi (g)$
for $g, h \in G, a \in \mathbb{R}$.  This will be useful in the sequel.

We seek an obstruction $\eta$ to writing $U_n \in A(n)$ as
$U_{n-1} \otimes R_z(\alpha)$, written in terms of characters.
First, let us classify which diagonal $U_n$ may be written in this
way.

\begin{proposition}[cf. {\cite[\S2.2]{BullockMarkov:02}}]
\label{prop:tensor}
Let $U = \sum_{j=0}^{N-1} u_j \ket{j} \bra{j}$.  Then
there exists $V=\sum_{j=0}^{N/2-1} v_j \ket{j} \bra{j}$
in $A(n-1)$ and
$W=w_0 \ket{0} \bra{0}+w_1 \ket{1}\bra{1}$ a one-qubit diagonal
so that $U=V \otimes W$ if and only if
\begin{equation}
u_0 u_1^{-1} = u_2 u_3^{-1} = u_4 u_5^{-1} = \cdots = u_{N-2} u_{N-1}^{-1}
\end{equation}
\end{proposition}

\begin{proof}
The check that such a tensor satisfies the chain of equalities is routine.
For the opposite implication, let $U = \sum_{j=0}^{N-1} u_j \ket{j} \bra{j}$.
Then define the $W=u_0 \ket{0} \bra{0}+u_1 \ket{1}\bra{1}$.
Now $U$ being unitary demands $u_0 \neq 0$.  Thus, choose in the expression
for $V$ that
$v_0=1,v_1=u_2/u_0,v_2=u_4/u_0,\cdots,v_{N/2-1}=u_{N-2}/u_0$.
The chain equality then implies $U=V\otimes W$.
\end{proof}

We now introduce the language for our obstruction $\eta$.  Note that
corollary \ref{cor:obstruction} motivates these
technical terms and is crucial to the constructions of Section \ref{sec:xr}
and Appendix \ref{sec:cr}.

\begin{definition}
\label{def:chi&eta}
Let $U=\sum_{j=0}^{N-1} u_j \ket{j} \bra{j}$ define coordinates on
$A(n)$.
For $1 \leq j \leq  N/2-1$, we define character functions
\mbox{$\chi_j :A(n) \rightarrow \mathbb{C}-\{0\}$} by
$\chi_j (U) = u_{2j-2} u_{2j-1}^{-1} u_{2j}^{-1} u_{2j+1}$.
For $U \in A(n)$, we define the vector valued function
$\eta : A(n) \rightarrow \mathbb{R}^{N/2-1}$
by $\eta (U) = -i \; [ \log \chi_1 (U) \ \log \ \chi_2 (U) \ \cdots
\ \log \chi_{N/2-1} (U) ]^t$.  Here, the superscript denotes the
transpose of the typeset row vector, so that we follow that
the usual convention of linear algebra that vector-valued functions
output column vectors.
\end{definition}

\begin{corollary}
\label{cor:obstruction}
The function $\eta : A(n) \rightarrow \mathbb{R}^{N/2-1}$
has the following properties.
\begin{itemize}
\item $[U=V \otimes W \mbox{ for }V \in A(n-1), W \in A(1)]
\Longleftrightarrow [\eta (U) = \vec{0}]$
\item For $U_1, U_2 \in A(n)$, we have
$\eta (U_1 U_2) = \eta (U_1) + \eta (U_2)$.
\item For $U \in A(n)$, $a \in \mathbb{R}$, we have
$\eta ( U^a) = a \; \eta (U)$.
\end{itemize}
\end{corollary}

Hence, the function $\eta (-)$ is a quantitative obstruction
to writing $U$ as a tensor on the last line.  A heuristic for the algorithms
of Sections \ref{sec:xr} and \ref{sec:cr} would then be the following.
\begin{enumerate}
\item  Define a large enough set of parameter dependent circuit blocks
in $A(n)$ so as to control all $N/2-1$ degrees of freedom of $\eta$.
Note this number of degrees of freedom coincides with the number of
nonempty subsets of the top lines $\{1,2,\cdots ,n-1\}$.
\item  Use the previous construction and the properties of $\eta$
to append circuit blocks to $U$ so that $\eta = \vec{0}$.
Then the composition $\tilde{U}=V \otimes W$, with $W$ some $R_z(\alpha)$
gate up to global phase.
\item  Recurse on $V$.
\end{enumerate}
In terms of this heuristic, the circuit blocks of Section \ref{sec:cr} are
the usual conditioned gates $\Lambda_k [R_z(\alpha)]$ \cite{BarencoEtAl:95},
while Section \ref{sec:xr}
requires a variant {\tt XOR}-controlled rotation.  We denote this
$\bigoplus_k [R_z(\alpha)]$, in analogy to the $\Lambda$ of
$\Lambda_k [V]$ being an enlarged version of the
propositional logic symbol $\wedge$ for ${\tt AND}$.

\section{Synthesis using $\bigoplus_k [R_z(\alpha)]$}
\label{sec:xr}

This section describes our main synthesis algorithm.  Certain proofs are
omitted due to their similarity to results of Appendix \ref{sec:cr}.
This appendix may be read first independently in order to motivate
the constructions in this section.

\subsection*{Circuit blocks for $\bigoplus_k [R_z(\alpha)]$}
\label{sec:subset}

We begin by making precise the notion of a $k$-fold {\tt XOR}-controlled
one-qubit computation $V \in U(2^1)$.  Several circuits blocks holding
$2k+1$ elementary gates are associated with this for
$V=R_z(\alpha)$.  Thus we first describe the $(k+1)$-qubit computation,
then highlight a circuit optimized for cancellation in our application,
and finally describe possible variant circuit blocks.

\begin{definition}  Let $k \geq 1$, $V \in U(2^1)$ a one-qubit
quantum computation, and for $b_1, b_2, \cdots, b_{k+1} \in \mathbb{F}_2$ let
the bit-string
$b_1 b_2 \cdots b_{k+1}$ also denote the element of $\mathbb{Z}$ with this
binary representation.  Then the {\tt XOR}-controlled $V$-computation
controlled on lines $1,2,\cdots,k$ and acting on line $k+1$ is that
$\bigoplus_k (V) \in U(2^{k+1})$ which extends linearly from
\begin{equation}
[{{\bigoplus}}_k (V)] \ket{b_1 b_2 \cdots b_{k+1}} =
\left\{
\begin{array}{l}
\ket{b_1 \cdots b_k} \otimes V\ket{b_{k+1}}, \mbox{ if} \\
 \quad b_1 \oplus b_2 \oplus
\cdots \oplus b_{k} = 0 \in \mathbb{F}_2 \\
\ket{b_1 b_2 \cdots b_k} \otimes {\bf V^*} \ket{b_{k+1}}, \mbox{ if} \\
\quad  b_1 \oplus b_2 \oplus
\cdots \oplus b_{k} = 1 \in \mathbb{F}_2 \\
\end{array}
\right.
\end{equation}
Here, ${\bf V^*} \in U(2^1)$ is the inverse or adjoint operator to $V$ and
the symbol $\oplus$ denotes the exclusive-{\tt OR} operation which is also
addition in $\mathbb{F}_2$.
We take the convention that $\bigoplus_0 (V) \ket{b_1 b_2 \cdots b_{n}}=
\ket{b_1 b_2 \cdots b_{n-1}} \otimes V\ket{b_{n}}$.
In $n$ qubits, should
$S \subset \{ 1,2,\cdots ,n-1\}$ be a possibly empty subset,
we write $\bigoplus_S (V)$ for the instance of
$\bigoplus_{\# S}(V)$ conditioned on lines $\{j \in S\}$ and
acting on line $n$.
\end{definition}

In the application, we will use the circuit diagram for
$\bigoplus_k [R_z(\alpha)]$ which follows from the following
equation.
Let $S \subset \{1,2,\cdots, n-1\}$, say nonempty, label
$S=\{s_1,s_2,\cdots, s_k\}$, and finally let ${\bf 1} \in
U(N/2)$ denote an $(n-1)$-qubit identity computation.
Recalling the controlled-not notation
$C_j^k$ from the Introduction one has
\begin{equation}
\label{eq:stdXR}
{\bigoplus}_S [R_z(\alpha)] =
C_{s_1}^n C_{s_2}^n \cdots C_{s_{k-1}}^n C_{s_k}^n
[{\bf 1} \otimes R_z(\alpha)] C_{s_k}^n C_{s_{k-1}}^n \cdots
C_{s_2}^n C_{s_1}^n
\end{equation}
All controlled-not gates to either side of the
${\bf 1} \otimes R_z(\alpha)$ term commute.
The right hand side of figure
\ref{fig:circS} illustrates the corresponding circuits.
These circuits require $2k+1$ elementary gates
and are the implementation of
$\bigoplus_k [R_z(\alpha)]$ used in our final circuit
diagrams.  For completeness,
we briefly note possible variant circuit blocks of the same size.

Let $S \subset \{1,\cdots, n-1\}$ and $S\neq\emptyset$.
Suppose $S = \{s_1, \cdots s_k\}$ with $s_1<s_2<\cdots < s_k$.
Then another quantum circuit for $\bigoplus_S [R_z(\alpha)]$
arises from
\begin{equation}
\label{eq:altXR}
{\bigoplus}_S[R_z(\alpha)]=
\begin{array}{l}
C_{s_1}^{s_2} C_{s_2}^{s_3} \cdots C_{s_{k-1}}^{s_k}
C_{s_k}^n \; \; [{\bf 1} \otimes R_z(\alpha)] \\
\quad
C_{s_k}^n C_{s_{k-1}}^{s_k} \cdots
C_{s_2}^{s_3} C_{s_1}^{s_2} \\
\end{array}
\end{equation}
This is illustrated to the left in Figure \ref{fig:circS}.

\begin{figure*}[t]
\begin{center}
\begin{picture}(51,7)
\put(0,0){\usebox{\hwire}} \put(1,3){\circle{1}}
\put(0,2){\usebox{\hwire}} \put(1,7){\circle*{.4}}
\put(1,7){\line(0,-1){4.5}} \put(0,4){\usebox{\hwire}}
\put(0,6){\usebox{\hwire}}

\put(1,0){\usebox{\topCNOT}} \put(1,2){\usebox{\hwire}}
\put(1,4){\usebox{\hwire}} \put(1,6){\usebox{\hwire}}

\put(3,0){\usebox{\Rzgate}} \put(3,2){\usebox{\hwire}}
\put(3,4){\usebox{\hwire}} \put(3,6){\usebox{\hwire}}

\put(5,0){\usebox{\topCNOT}} \put(5,2){\usebox{\hwire}}
\put(5,4){\usebox{\hwire}} \put(5,6){\usebox{\hwire}}

\put(6,0){\usebox{\hwire}} \put(7,3){\circle{1}}
\put(6,2){\usebox{\hwire}} \put(7,7){\circle*{.4}}
\put(7,7){\line(0,-1){4.5}} \put(6,4){\usebox{\hwire}}
\put(6,6){\usebox{\hwire}}

\put(9,4){\makebox(0,0)[bl]{\LARGE $=$}}

\put(11,0){\usebox{\hwire}} \put(12,3){\circle*{0.4}}
\put(12,7){\circle{1}} \put(12,3){\line(0,1){4.5}}
\put(11,2){\usebox{\hwire}} \put(11,4){\usebox{\hwire}}
\put(11,6){\usebox{\hwire}}

\put(13,1){\circle{1}} \put(13,7){\circle*{0.4}}
\put(13,7){\line(0,-1){6.5}} \put(12,0){\usebox{\hwire}}
\put(12,2){\usebox{\hwire}} \put(12,4){\usebox{\hwire}}
\put(12,6){\usebox{\hwire}}

\put(14,0){\usebox{\Rzgate}} \put(14,2){\usebox{\hwire}}
\put(14,4){\usebox{\hwire}} \put(14,6){\usebox{\hwire}}

\put(17,1){\circle{1}} \put(17,7){\circle*{0.4}}
\put(17,7){\line(0,-1){6.5}} \put(16,0){\usebox{\hwire}}
\put(16,2){\usebox{\hwire}} \put(16,4){\usebox{\hwire}}
\put(16,6){\usebox{\hwire}}

\put(17,0){\usebox{\hwire}} \put(18,3){\circle*{0.4}}
\put(18,7){\circle{1}} \put(18,3){\line(0,1){4.5}}
\put(17,2){\usebox{\hwire}} \put(17,4){\usebox{\hwire}}
\put(17,6){\usebox{\hwire}}

\put(20,4){\makebox(0,0)[bl]{\LARGE $=$}}

\put(22,1){\line(1,0){1.0}} \put(22,3){\line(1,0){1.0}}
\put(22,5){\line(1,0){1.0}} \put(22,7){\line(1,0){1.0}}

\put(23,0){\usebox{\Rzgate}} \put(23,2){\usebox{\hwire}}
\put(23,4){\usebox{\hwire}} \put(23,6){\usebox{\hwire}}
\put(24,3){\circle*{0.4}} \put(24,5){\circle*{0.4}}
\put(24,7){\circle*{0.4}}

\put(24,4){\circle{0.8}} \put(24,6){\circle{0.8}}
\put(23.6,4){\line(1,0){0.8}} \put(23.6,6){\line(1,0){0.8}}
\put(24,1.8){\line(0,1){5.2}}

\put(25,1){\line(1,0){1.0}} \put(25,3){\line(1,0){1.0}}
\put(25,5){\line(1,0){1.0}} \put(25,7){\line(1,0){1.0}}

\put(27,4){\makebox(0,0)[bl]{\LARGE $=$}}

\put(29,0){\usebox{\hwire}} \put(30,1){\circle{1}}
\put(29,2){\usebox{\hwire}} \put(30,7){\circle*{.4}}
 \put(30,7){\line(0,-1){6.5}}
  \put(29,4){\usebox{\hwire}}
\put(29,6){\usebox{\hwire}}

\put(31,0){\usebox{\topCNOT}} \put(31,2){\usebox{\hwire}}
\put(31,4){\usebox{\hwire}} \put(31,6){\usebox{\hwire}}

\put(33,0){\usebox{\Rzgate}} \put(33,2){\usebox{\hwire}}
\put(33,4){\usebox{\hwire}} \put(33,6){\usebox{\hwire}}

\put(35,0){\usebox{\topCNOT}} \put(35,2){\usebox{\hwire}}
\put(35,4){\usebox{\hwire}} \put(35,6){\usebox{\hwire}}

\put(37,0){\usebox{\hwire}} \put(38,1){\circle{1}}
\put(37,2){\usebox{\hwire}} \put(38,7){\circle*{.4}}
\put(38,7){\line(0,-1){6.5}} \put(37,4){\usebox{\hwire}}
\put(37,6){\usebox{\hwire}}

\put(39,4){\makebox(0,0)[bl]{\LARGE $=$}}

\put(41,0){\usebox{\hwire}} \put(44,1){\circle{1}}
\put(41,2){\usebox{\hwire}} \put(44,7){\circle*{.4}}
 \put(44,7){\line(0,-1){6.5}}
 \put(41,4){\usebox{\hwire}}
\put(41,6){\usebox{\hwire}}

\put(41,0){\usebox{\topCNOT}} \put(43,0){\usebox{\hwire}}
                              \put(43,2){\usebox{\hwire}}
\put(43,4){\usebox{\hwire}} \put(43,6){\usebox{\hwire}}

\put(45,0){\usebox{\Rzgate}} \put(45,2){\usebox{\hwire}}
\put(45,4){\usebox{\hwire}} \put(45,6){\usebox{\hwire}}

\put(47,0){\usebox{\topCNOT}} \put(47,2){\usebox{\hwire}}
\put(47,4){\usebox{\hwire}} \put(47,6){\usebox{\hwire}}

\put(49,0){\usebox{\hwire}} \put(50,1){\circle{1}}
\put(49,2){\usebox{\hwire}} \put(50,7){\circle*{.4}}
\put(50,7){\line(0,-1){6.5}} \put(49,4){\usebox{\hwire}}
\put(49,6){\usebox{\hwire}}

\end{picture}
\begin{caption}
{ \label{fig:circS}
Shown at center is a symbol due to the authors for
denoting {\tt XOR} control.  At right are circuits
for $\bigoplus_S [R_z(\alpha)]$ per Equation
\ref{eq:stdXR}, as used in the circuits diagonal computations.
Here, $n=4$ qubits and $S=\{1,3\}\subset \{1,2,3\}$, so this is
an instance of $\bigoplus_2 [R_z(\alpha)]$.
At left are possible variaint circuits per Equations
\ref{eq:altXR} and \ref{eq:XRsigma}, where
$\sigma$ is an identity permutation and
$\sigma$ is the flip permutation of two elements.
}
\end{caption}
\end{center}
\end{figure*}

Finally, although the controlled-not gates in the second
diagram corresponding to the alternate Equation \label{eq:altXRcont}
certainly do \emph{not} commute, one may reorder the circuit in
a certain sense.
Let $\sigma$ be a permutation of $\{1,\cdots,k\}$, retaining
$S=\{s_1 < s_2 < \cdots < s_k\}$.
\begin{equation}
\label{eq:XRsigma}
{\bigoplus}_S [R_z(\alpha)]=
\begin{array}{c}
C_{s_{\sigma(1)}}^{s_{\sigma(2)}}
C_{s_{\sigma(2)}}^{s_{\sigma(3)}}
\cdots C_{s_{\sigma(k-1)}}^{s_{\sigma(k)}}
C_{s_\sigma(k)}^n
\; \; [{\bf 1} \otimes R_z(\alpha)]
\\
C_{s_{\sigma(k)}}^n C_{s_{\sigma(k-1)}}^{s_{\sigma(k)}}
\cdots C_{s_{\sigma(2)}}^{s_{\sigma(3)}}
C_{s_{\sigma(1)}}^{s_{\sigma(2)}} \\
\end{array}
\end{equation}
See the left hand side of Figure \ref{fig:circS}.

\subsection*{Computation of $\eta(\; \bigoplus_S[R_z(\alpha)] \; )$}

We find it more convenient to use mathematical notation for vectors such
as values of $\eta$ rather than the {\tt bra-ket} notation.  We briefly
recall the appropriate conventions, treated in more detail in
Appendix \ref{sec:cr}.

\begin{definition}
\label{def:vdef}
For $1\leq j \leq N/2-1$, let $e_j$ denote is the column
vector in $\mathbb{R}^{N/2-1}$ with
a single entry of $1$ in the $j^{\mbox{th}}$
row and all other entries $0$.
The vectors $v_j=e_j-e_{j+1}$ if $1 \leq j \leq N/2-2$,
while $v_0=-e_1$ and $v_{N/2-1}=e_{N/2-1}$.
\end{definition}

The vectors $\{v_j \; ; 1 \leq j \leq N/2-1\}$ form a basis for
$\mathbb{R}^{2^{n-1}-1}$.  We need one more definition before
computing $\eta( \; \oplus_S [R_z(\alpha)] \; )$.

\begin{definition} Let
$S=\{s_1,s_2,\cdots,s_k\} \subset \{1,2,\cdots, n-1\}$ be nonempty.
In $n$ qubits with $N=2^n$, let $1 \leq j \leq N/2-1$ with
binary representation $j=b_1b_2 \cdots b_{n-1}$ for
$b_1,b_2,\cdots, b_{n-1} \in \mathbb{F}_2$.  Then we say
the integer $j$ is {\tt XOR}-$S$-conditioned iff
\mbox{$b_{s_1}\oplus b_{s_2} \oplus \cdots \oplus b_{s_k}=1$}.
We further define the set
\begin{equation}
\mathcal{F}(S) =
\{ 1 \leq j \leq N/2-1 \; ; \; j \mbox{ is {\tt XOR}-$S$-conditioned}\}
\end{equation}
By a \emph{flip state} of $S$, we mean any $j \in \mathcal{F}(S)$, i.e.,
$S$-flip is an abbreviation of {\tt XOR}-$S$-conditioned.
\end{definition}

\begin{example}
\label{ex:4flips}
Consider the special case of $n=4$ qubits.  The flip
states of each nonempty
subset of $\{1,2,3\}$ of the top three lines are given
in the table below, in binary.
\begin{center}
\begin{tabular}{|c||c|}
\hline
subset & flip states \\
\hline
\hline
$\{1\}$ & $100$, $101$, $110$, $111$ \\
\hline
$\{1,2\}$ & $010$, $011$, $100$, $101$ \\
\hline
$\{1,3\}$ & $001$, $011$, $100$, $110$ \\
\hline
$\{1,2,3\}$ & $001$, $010$, $100$, $111$ \\
\hline
$\{2\}$ & $010$, $011$, $110$, $111$ \\
\hline
$\{2,3\}$ & $001$, $010$, $101$, $110$ \\
\hline
$\{3\}$ & $001$, $011$, $101$, $111$ \\
\hline
\end{tabular}
\end{center}
Note that for any $S \neq \emptyset$, exactly half of the eight
integers $0,1,\cdots,7$ are elements of $\mathcal{F}(S)$.
\end{example}

\begin{proposition}
\label{prop:vecviaflips}
Let $\mathcal{F}(S)$ be the set of flip states of any nonempty
$S \subset \{1,2,\cdots,n-1\}$.  Then
\begin{equation}
\label{eq:flipvec}
\eta(\; {\bigoplus}_S[R_z(\alpha)] \;) =
-2 \alpha \sum_{j \in \mathcal{F}(S)} v_j
\end{equation}
Also, for $S=\emptyset$, $\eta[{\bf 1} \otimes R_z(\alpha)]=\vec{0}$.
\end{proposition}

The proof is similar to that of Proposition \ref{prop:LogChiCR}.  However,
$\oplus_S [R_z(\alpha)]$ never leaves any computational basis state fixed,
which accounts for the factor of two.

\begin{example}
\label{ex:computevec}
Consider $n=4$ qubits for the subset $S=\{1,3\}$ and $\alpha$ arbitrary.
For convenience, label $\phi = -\alpha/2$, so that
$R_z(\alpha) = \mbox{e}^{i \phi} \ket{0}\bra{0}
+ \mbox{e}^{-i \phi} \ket{1}\bra{1}$.  We
leave it to the reader to check that $V=\bigoplus_S[ R_z(\alpha)]$ is
diagonal and merely describe the multiples on each computational basis state.
\begin{center}
\small
\begin{tabular}{|l|c||l|c||l|c||l|c|}
\hline
state & mult & state & mult & state & mult & state & mult \\
\hline
\hline
$\ket{0000}$ & $\mbox{e}^{i\phi}$ & $\ket{0100}$ & $\mbox{e}^{i\phi}$ &
$\ket{1000}$ & $\mbox{e}^{-i\phi}$ & $\ket{1100}$ & $\mbox{e}^{-i\phi}$ \\
\hline
$\ket{0001}$ & $\mbox{e}^{-i\phi}$ & $\ket{0101}$ & $\mbox{e}^{-i\phi}$ &
$\ket{1001}$ & $\mbox{e}^{i\phi}$ & $\ket{1101}$ & $\mbox{e}^{i\phi}$ \\
\hline
$\ket{0010}$ & $\mbox{e}^{-i\phi}$ & $\ket{0110}$ & $\mbox{e}^{-i\phi}$ &
$\ket{1010}$ & $\mbox{e}^{i\phi}$ & $\ket{1110}$ & $\mbox{e}^{i\phi}$ \\
\hline
$\ket{0011}$ & $\mbox{e}^{i\phi}$ & $\ket{0111}$ & $\mbox{e}^{i\phi}$ &
$\ket{1011}$ & $\mbox{e}^{-i\phi}$ & $\ket{1111}$ & $\mbox{e}^{-i\phi}$ \\
\hline
\end{tabular}
\end{center}
Thus, $\chi_1(V)= \mbox{e}^{4i\phi}$,
$\chi_2(V)=\mbox{e}^{-4i\phi}$, $\chi_3(V)=\mbox{e}^{4i\phi}$,
$\chi_4(V)=1$, $\chi_5(V)=\mbox{e}^{-4i\phi}$,
$\chi_6(V)=\mbox{e}^{4i\phi}$, and
$\chi_7(V)=\mbox{e}^{-4i\phi}$.  Thus we have
computed
$\eta (\; \bigoplus_{\{1,3\}}[R_z(\alpha)] \;) =
4 \phi i [1 \ -1 \ 1 \ 0 \ -1 \ 1 \ -1]^t$.

On the other hand, flip states for $\{1,3\}$ are given in binary by $j=
001, 011, 100,$ and $110$.  So $\mathcal{F}(S)=\{1,3,4,6\}$ and
\[
(e_1-e_2)+(e_3-e_4)+(e_4-e_5)+(e_6-e_7)=[1 \ -1 \ 1 \ 0 \ -1 \ 1 \ -1]^t.
\]
This concludes the example.
\end{example}

\subsection*{$\bigoplus_k [R_z(\alpha)]$-block synthesis algorithm}
\label{sec:XRalgo}

The $-0.5$ radians in the Definition of the following matrix
cancels the $2$ coefficient in Equation \ref{eq:flipvec}, so
that the resulting matrix has all entires in $\mathbb{Z}$.  It is similar
to Definition \ref{def:LogCRMat}.

\begin{definition}
\label{def:LogXRMat}
The matrix ${\bf \eta}^\oplus$
is the $(N/2-1) \times (N/2-1)$ real matrix defined as follows.
Order nonempty subsets
$S_1$, $S_2$, $\cdots$, $S_{N/2-1}$ in \emph{Grey order},
omitting the empty set.  Then for
$1 \leq j \leq N/2-1$, the $j^{\mbox{th}}$ column of
${\bf \eta}^\oplus$ is
$\eta( \; \bigoplus_{S_j} [R_z(-0.5\mbox{ radians})] \;)$.
\end{definition}

\begin{example}
\label{ex:CHI4}
Computing the four-qubit case of ${\bf \eta}^\oplus$ is most quickly
accomplished using the table of example \ref{ex:4flips} and
Proposition \ref{prop:vecviaflips}.
The Grey order of nonempty subsets of $\{1,2,3\}$ is
$\{3\}$,$\{2,3\}$,$\{2\}$,$\{1,2\}$,$\{1,2,3\}$,$\{1,3\}$,
$\{1\}$.  Thus the Definition in this case states
\begin{equation}
{\bf \eta}^\oplus =
\left(
\begin{array}{rrrrrrr}
1 & 1 & 0 & 0 & 1 & 1 & 0 \\
-1 & 0 & 1 & 1 & -1 & 0 & 0 \\
1 & -1 & 0 & 0 & 1 & -1 & 0 \\
-1 & 0 & -1 & 0 & 0 & 1 & 1 \\
1 & 1 & 0 & 0 & -1 & -1 & 0 \\
-1 & 0 & 1 & -1 & 1 & 0 & 0  \\
1 & -1 & 0 & 0 & -1 & 1 & 0 \\
\end{array}
\right)
\end{equation}
The fifth column recalls example \ref{ex:computevec}.
\end{example}

The matrix ${\bf \eta}^\oplus$ has the following application.
Note the right hand side is matrix multiplication with the
column vector $\vec{\alpha}$.

\begin{lemma}
\label{lem:XRcomp}
Fix $n$ qubits, with $N=2^n$.
Let $\vec{\alpha}=[\alpha_1 \ \cdots \ \alpha_{N/2-1}]^t$ be a
vector of angles, $0 \leq \alpha_j < 2\pi$, $1 \leq j \leq N/2-1$.
Then for $S_1$, $S_2$, \dots $S_{N/2-1}$ the Grey ordering
of the nonempty subsets of the set of top lines $\{1,\cdots,n-1\}$,
\begin{equation}
\eta( \; \oplus_{S_1} [R_z(\alpha_1)] \; \cdots
\oplus_{S_{N/2-1}} [R_z(\alpha_{N/2-1})] \; ) \; \; = \; \;
-2 \; {\eta}^\oplus \vec{\alpha}
\end{equation}
\end{lemma}

The proof is quite similar to Lemma \ref{lem:CRcomp}.  It uses
Proposition \ref{prop:vecviaflips} and properties of $\eta(-)$
following from each component being a character.

We now state the synthesis algorithm.  It is critical in the following
that ${\bf \eta}^\oplus$ be invertible.  This result will be proven in
the next subsection.

\noindent
{\bf {\tt XOR}-Controlled Rotation Synthesis Algorithm}
Let $U=\sum_{j=0}^{N-1} u_j \ket{j} \bra{j}$, for which
we wish to synthesize a circuit diagram using
$\bigoplus_k [R_z(\alpha)]$ blocks.
Label $S_1$, $S_2$, $S_3$ \dots $S_{N/2-1}$ the
nonempty subsets of the top lines $\{1,\cdots, n-1\}$ in the
\emph{Grey order}.
\begin{enumerate}
\item{ Compute $\vec{\psi}= \eta(U)$.}
\item{ Compute the inverse matrix $({\bf \eta}^\oplus)^{-1}$.}
\item{ Compute $\vec{\alpha}=(-1/2) ({\bf \eta}^\oplus)^{-1} \vec{\psi}$,
treating $\vec{\psi}$ as a column vector.  Label
$\vec{\alpha}=[\alpha_1 \  \cdots \alpha_{N/2-1}]^t$.}
\item{ Compute the diagonal quantum computation
\(
\tilde{U}
= {\bigoplus}_{S_1}[R_z(-\alpha_1)] \; \cdots \;
{\bigoplus}_{S_{N/2-1}}[R_z (-\alpha_{N/2-1})] \; U
\)
As is verified below, $\tilde{U}$ is a tensor.}
\item{ Use the argument of Proposition \ref{prop:tensor} to compute
$\tilde{U}=V \otimes W$ for $V \in A(n-1)$ and
$W=\mbox{e}^{i \Phi}R_z(\alpha_0)$ for some angle $\alpha_0$.}
\item{ Given prior computations, the following
expression holds:
\begin{equation}
\begin{array}{lcl}
U & = & {\bigoplus}_{\emptyset}[R_z(\alpha_0)] \;
{\bigoplus}_{S_1}[R_z(\alpha_1)] \;
\cdots \\
& &  \quad {\bigoplus}_{S_{N/2-1}}[R_z(\alpha_{N/2-1})] \; [V \otimes {\bf 1}]
\\
\end{array}
\end{equation}
Here, ${\bf 1}$ denotes the trivial computation of $U(2^1)$.
Also, ${\bigoplus}_\emptyset [R_z(\alpha_0)]$ means ${\bf 1} \otimes
R_z(\alpha_0)$ for ${\bf 1} \in U(2^{N/2})$.}
\item
Decompose each ${\bigoplus}_k R_z(\alpha)$ into elementary gates using the
circuit diagrams at the right of Figure \ref{fig:circS}.
\item  Using the Grey order
and $C^n_{j} C^n_k = C^n_k C^n_j$,
cancel all but one controlled-not between consecutive $R_z(\alpha)$ gates
in the resutling diagram.
\item{
The algorithm terminates by recursively producing a circuit diagram for
$V \in A(n-1)$.}
\end{enumerate}

\begin{example}
Consider the following $3$-qubit computation:
\begin{equation}
\begin{array}{lcl}
U & = &
\mbox{e}^{4\pi i/12} \ket{0}\bra{0} +
\mbox{e}^{2\pi i/12} \ket{1}\bra{1} + \\
& &
\mbox{e}^{9\pi i/12} \ket{2}\bra{2} +
\mbox{e}^{7\pi i/12} \ket{3}\bra{3} + \\
& & \mbox{e}^{3\pi i/12} \ket{4}\bra{4} +
\mbox{e}^{8\pi i/12} \ket{5}\bra{5} + \\
& &
\mbox{e}^{11\pi i/12} \ket{6}\bra{6} +
\mbox{e}^{10\pi i/12}) \ket{7}\bra{7}
\end{array}
\end{equation}
We apply the synthesis algorithm above to $U$.

We begin by computing the $3$-qubit case of ${\bf \eta}^\oplus$.
The Grey order is $\{1\}$, $\{1,2\}$, and $\{2\}$.
\begin{equation}
{\eta}^\oplus =
\left(
\begin{array}{rrr}
1 & 1 & 0 \\
-1 & 0 & 1 \\
1 & -1 & 0 \\
\end{array}
\right)
\end{equation}
The inverse matrix appears in the algorithm and may be reused for
multiple diagonal computations.
\begin{equation}
({\bf \eta}^\oplus)^{-1} =
(1/2)
\left(
\begin{array}{rrr}
1 & 0 & 1 \\
1 & 0 & -1 \\
1 & 2 & 1 \\
\end{array}
\right)
\end{equation}
Now $\vec{\psi}= \eta(U)=
-i [\log \chi_1(U) \  \log \chi_2(U) \  \log \chi_3(U)]^t
=[0  \ 7\pi/12 \ -6 \pi/12]^t$.  Thus computing
the parameters for the ${\bigoplus}_S[R_z(\alpha)]$ blocks,
\begin{equation}
\vec{\alpha}=(-1/2)({\bf\eta}^\oplus)^{-1} \vec{\psi}=
[3 \pi/24 \ \ -3\pi/24 \ \ -4\pi/24]^t
\end{equation}
It should be the case that the computation $\tilde{U}$ given by
\begin{equation}
{\bigoplus}_{\{1\}}[R_z(-3\pi/24)]
\; {\bigoplus}_{\{1,2\}}[R_z(3\pi/24)] \;
{\bigoplus}_{\{2\}}[R_z(4\pi/24)] \; U
\end{equation}
has $\tilde{U}=V\otimes W$ for $V$ a two-qubit diagonal and $W$ a one-qubit
diagonal.  We verify this by computing matrix coefficients for
$\tilde{U}$.

In the following computation, for given $R \in A$ we abbreviate
$R = \sum_{j=0}^{N-1} r_j \ket{j} \bra{j}$ as
$R=\mbox{diag}(r_0,r_1,\cdots,r_{N-1})$ in order to save space.
The first step in computing $\tilde{U}$
is to compute $\bigoplus_{\{1\}}[R_z(4\pi/24)]$.
Begin by noting that
\begin{equation}
\begin{array}{lcl}
{\bf 1} \otimes {\bf 1} \otimes R_z(4\pi/24) & = & \mbox{diag}
(
\mbox{e}^{-4\pi i/48},\mbox{e}^{4\pi i/48},
\mbox{e}^{-4\pi i/48},\mbox{e}^{4\pi i/48}, \\
& & \mbox{\quad \quad}
\mbox{e}^{-4\pi i/48},\mbox{e}^{4\pi i/48},
\mbox{e}^{-4\pi i/48},\mbox{e}^{4\pi i/48}) \\
\end{array}
\end{equation}
Associating the entries with $\ket{000}$, $\ket{001}$, etc., we reverse
those pairs $\ket{b_1 b_2 b_3}$ with the binary integer
$b_1 b_2 \in \mathcal{F}(\{1\})$.
\begin{equation}
\begin{array}{lcl}
{\bigoplus}_{\{1\}}[R_z(4\pi/24)] & = & \mbox{diag}
(
\mbox{e}^{-4\pi i/48},\mbox{e}^{4\pi i/48},
\mbox{e}^{-4\pi i/48},\mbox{e}^{4\pi i/48}, \\
& & \mbox{\quad \quad}
\mbox{e}^{4\pi i/48},\mbox{e}^{-4\pi i/48},
\mbox{e}^{4\pi i/48},\mbox{e}^{-4\pi i/48}) \\
\end{array}
\end{equation}
We may similarly construct $\bigoplus_{\{1,2\}}[R_z(3\pi/24)]$.
\begin{equation}
\begin{array}{lcl}
{\bigoplus}_{\{1,2\}}[R_z(3\pi/24)] & = & \mbox{diag}
(
\mbox{e}^{-3\pi i/48},\mbox{e}^{3\pi i/48},
\mbox{e}^{3\pi i/48},\mbox{e}^{-3\pi i/48}, \\
& & \mbox{\quad \quad}
\mbox{e}^{3\pi i/48},\mbox{e}^{-3\pi i/48},
\mbox{e}^{-3\pi i/48},\mbox{e}^{3\pi i/48}) \\
\end{array}
\end{equation}
Finally, the flip states of $\{2\}$ are $j=1,3$.  Thus
\begin{equation}
\begin{array}{lcl}
{\bigoplus}_{\{2\}}[R_(-3\pi/24)] & = &
\mbox{diag} (
\mbox{e}^{3\pi i/48},\mbox{e}^{-3\pi i/48},
\mbox{e}^{-3\pi i/48},\mbox{e}^{3\pi i/48}, \\
& & \mbox{\quad \quad}
\mbox{e}^{3\pi i/48},\mbox{e}^{-3\pi i/48},
\mbox{e}^{-3\pi i/48},\mbox{e}^{3\pi i/48}) \\
\end{array}
\end{equation}
Collecting all terms, we arrive at
\begin{equation}
\begin{array}{llll}
\tilde{U} & = &
\mbox{diag}
(
\mbox{e}^{-4\pi i/48},\mbox{e}^{4\pi i/48},
\mbox{e}^{-4\pi i/48},\mbox{e}^{4\pi i/48}, \\
& & \mbox{\quad \quad}
\mbox{e}^{4\pi i/48},\mbox{e}^{-4\pi i/48},
\mbox{e}^{4\pi i/48},\mbox{e}^{-4\pi i/48}
) \circ \\
& & \mbox{diag}
(
\mbox{e}^{-3\pi i/48},\mbox{e}^{3\pi i/48},
\mbox{e}^{3\pi i/48},\mbox{e}^{-3\pi i/48}, \\
& & \mbox{\quad \quad}
\mbox{e}^{3\pi i/48},\mbox{e}^{-3\pi i/48},
\mbox{e}^{-3\pi i/48},\mbox{e}^{3\pi i/48}
) \circ \\
& & \mbox{diag}
(
\mbox{e}^{3\pi i/48},\mbox{e}^{-3\pi i/48},
\mbox{e}^{-3\pi i/48},\mbox{e}^{3\pi i/48}, \\
& & \mbox{\quad \quad}
\mbox{e}^{3\pi i/48},\mbox{e}^{-3\pi i/48},
\mbox{e}^{-3\pi i/48},\mbox{e}^{3\pi i/48}
) \circ \\
& & \mbox{diag}( \mbox{e}^{4\pi i/12},
\mbox{e}^{8\pi i/48},
\mbox{e}^{36\pi i/48},
\mbox{e}^{28\pi i/48}, \\
& & \mbox{\quad \quad}
\mbox{e}^{12\pi i/48},
\mbox{e}^{32\pi i/48},
\mbox{e}^{44\pi i/48},
\mbox{e}^{40\pi i/48}) \\
& = & \mbox{diag}(
\mbox{e}^{12 \pi i/48}, \mbox{e}^{12 i\pi/48},
\mbox{e}^{32 \pi i/48}, \mbox{e}^{32 i\pi/48}, \\
& & \mbox{\quad \quad}
\mbox{e}^{22 \pi i/48}, \mbox{e}^{22 i\pi/48},
\mbox{e}^{42 \pi i/48}, \mbox{e}^{42 i\pi/48}) \\
\end{array}
\end{equation}
Thus $\tilde{U}= \mbox{diag}(\mbox{e}^{12\pi i/48},\mbox{e}^{32 \pi i/48},
\mbox{e}^{22 \pi i/48}, \mbox{e}^{42 \pi i/48}) \otimes \mbox{diag}(1,1)$.
The odd happenstance that the latter tensor factor is an identity saves
one gate.

Next, write out circuit diagrams for each $\bigoplus_S [R_z(\alpha)]$
per the right hand side of Figure \ref{fig:circS}.  Since the chose
the Grey order $\{1\}, \{1,2\}$, $\{2\}$, cancelling controlled
not gates produces the leftmost $8$ elementary gates of figure
\ref{fig:3qcase}.  Finally, call the algorithm recursively on $V$.
The two-qubit case coincides with other work \cite[\S 2.2]{BullockMarkov:02}.
\end{example}

\subsection*{Proof of Correctness}
\label{sec:XRcorrect}

We briefly verify that $\tilde{U}=V \otimes W$.  First use
Proposition \ref{prop:vecviaflips} for
\begin{equation}
\eta(\; {\bigoplus}_{S_1}[R_z(-\alpha_1)] \; \cdots \;
{\bigoplus}_{S_{N/2-1}}[R_z(-\alpha_{N/2-1})] \; ) = 2 {\bf \eta}^\oplus
\vec{\alpha}
\end{equation}
Now by definition $\vec{\alpha}=(-1/2) ({\bf \eta}^\oplus)^{-1} \vec{\psi}$,
so that $2 {\bf \eta}^\oplus \alpha = -\vec{\psi}$.
\begin{equation}
\eta(\; {\bigoplus}_{S_1}[R_z(-\alpha_1)] \; \cdots \;
{\bigoplus}_{S_{N/2-1}}[R_z(-\alpha_{N/2-1})] \; ) = -\vec{\psi}
\end{equation}
Then the property $\eta(U_1 U_2) = \eta(U_1) + \eta(U_2)$ demands
\begin{equation}
\eta( \; {\bigoplus}_{S_1}[R_z(-\alpha_1)] \; \cdots
{\bigoplus}_{S_{N/2-1}}[R_z(-\alpha_{N/2-1})] \; U \; ) =
-\vec{\psi} + \vec{\psi} = \vec{0}
\end{equation}
So by the restatement of Proposition \ref{prop:tensor}, we have
$\tilde{U} = V \otimes W$.

There is one remaining unjustified (subtle) statement to check.
\begin{proposition}
\label{prop:independence}
${\bf \eta}^\oplus$ is an invertible
$(N/2-1)\times (N/2-1)$ real matrix for $n\geq 1$.
\end{proposition}

\begin{sketch}
It is equivalent to consider the question for an alternate basis of
$\mathbb{R}^{N/2-1}$.  Thus, choose instead the vectors
$\{v_j \; ; \; 1 \leq j \leq \mathbb{R}^{N/2-1} \}$ of
Definition \ref{def:vdef}.
In this alternate basis, the similar matrix $M$ corresponding to
${\bf \eta}^\oplus$ has an entry of $1$ for the $v_{j}$ component
whenever $j$ is a flip state for the $j^{\mbox{th}}$-set in Grey order.

Fix an nonempty subset $S$ of $\{1,2,\cdots,n-1\}$,
thus fixing a column of
${\bf \eta}^\oplus$.  We first claim there
precisely $2^{n-2}$ flip states for $S$.
To see this, observe that the equation $\oplus_{k\in S} b_k=1$ satisfied by
$S$-flip states defines an affine linear $\mathbb{F}_2$ subspace of
the finite-dimensional vector space $(\mathbb{F}_2)^{2^{n-1}}$.  Then this
number of elements corresponds to the dimension count, since any $\ell$
dimensional vector space with
$\mathbb{F}_2$-scalars must contain $2^\ell$ elements.

Next, fix $S_1 \neq S_2$ distinct nonempty subsets.  Then
the associated columns of $M$ share precisely $2^{n-3}$ positions in which
each has a nonzero, unit entry.  This is again a dimension count.
Note that since $S$-flip states satisfy $\oplus_{k\in S} b_k=1$,
$S_1 \neq S_2$.  Thus the codimension one subspaces corresponding to $S_1$
and $S_2$ intersect transversally in a codimension two subspace.

Given these claims, label $M = (m_{jk})$ and recall $\delta_j^k$ the
Kronecker delta which is $1$ for $j=k$ and zero else.  Now considerations
of the last two paragraphs demand that for the transpose (real adjoint)
$M^t$, \
$M^t M = (m_{kj}) (m_{j\ell}) = 2^{n-2}( \delta_j^\ell + 1)$.  An omitted
argument then shows $0 \neq \mbox{det}(M^t M)$, demanding
$(\mbox{det} \; M)^2 \neq 0$.  As $M$ is invertible and ${\eta}^\oplus$ is
similar to $M$, we must have ${\bf \eta}^\oplus$ invertible.
\end{sketch}

\subsection*{Gate Counts}
\label{sec:counts}

Our circuit diagrams are built from blocks realizing
$\bigoplus_S [R_z(\alpha)]$ at the right of Figure \ref{fig:circS},
and the choice of subsets in the Grey order causes a large cancellation of
controlled-not gates which is required for the $O(2^n)$
asymptotic. We now justify the gate count of $2^{n+1}-3$,
which for $n=2$ coincides with $5$ gates
\cite[\S2.2]{BullockMarkov:02}.

Except for the recursive call to $V$, the synthesis algorithm
writes elementary gates realizing the following computition.
\begin{equation}
{\bigoplus}_{\emptyset} [R_z(\alpha_0)] \;
{\bigoplus}_{S_1} [R_z(\alpha_1)] \; \cdots
{\bigoplus}_{S_{N/2-1}} [R_z(\alpha_{N/2-1})]
\end{equation}
Here, $\bigoplus_{\emptyset}[R_z(\alpha_0)]=(\mbox{e}^{-i\Phi})
({\bf 1} \otimes W)$ is the one-qubit gate resulting on the
last tensor factor due to zeroing the obstruction $\eta(-)$.  We
have used the commutativity of $A(n)$ to move this computation
to the front to preserve the full Grey order including $\emptyset$.

Now realize each of the $\bigoplus_S [R_z(\alpha)]$ blocks using
the circuits at the right of Figure \ref{fig:circS}.  Due to the
Grey order, all but one controlled-not gate will cancel between
any two consecutive $R_z$ gates on the bottom line.  Thus the gate
count in terms of elementary gates from the Introduction
should account for the following.
\begin{itemize}
\item  $2^{n-1}$ controlled rotations $R_z$, since this is the number
of possibly empty subsets of $\{1,2,\cdots,n-1\}$.
\item  $2^{n-1}$ controlled-not gates, since one lies to the right
of each $R_z$ gate.
\end{itemize}
Thus prior to the recursive call, in $n \geq 2$ qubits the algorithm
will place $2^n$ elementary gates.

To obtain the exact count, stop the recursive count at $n=2$ qubits.
\begin{equation}
2^n + 2^{n-1} + \cdots + 8 + 4 = 2^{n+1}-4
\end{equation}
The end case of recursion is for $n=1$.  Since any one-qubit diagonal
may be written $\mbox{e}^{i\Phi} R_z(\alpha)$, the remaining one-qubit
diagonal requires one elementary gate.  Thus the grand total is $2^{n+1}-3$
elementary gates.

\section{Stable Lower Bounds}
\label{sec:lower}

The section justifies the claim of stably-asymptotical optimality
in Theorem \ref{thm} using an argument similar to one by E. Knill
\cite[Theorem 3.4]{Knill:95}. We provide a greater level of detail and
tailor the discussion to synthesis within a subgroup $H \subset U(N)$.
Our argument is what simpler because we are dealing with
{\em elementary} gates from the Introduction while
Knill uses {\em basic} gates \cite{BarencoEtAl:95}.

Thus let $S \subset U(N)$.  We introduce the following convention:
\begin{equation}
\tilde{S} = \{ \mbox{e}^{i \Phi} V \; ; \; 0 \leq \Phi < 2\pi, V \in S \}
\end{equation}
This will allow us to ignore global phases in the following discussion.
Note that $\tilde{A}=A$.

We now expand on comments made briefly in Definition
\ref{def:stablyasymptoticallyoptimal} of the Introduction.
A \emph{circuit topology}\footnote{We discuss here circuit topologies
in the elementary gate library.} $\tau$ is
an $n$-line diagram on which is marked a sequence
of gate-holders.  These gate-holders
are either controlled-not gates joining any two lines
or boxes labelled either $Y$ or $Z$.  To specialize the circuit topology
$\tau$ to an actual circuit, one chooses paramaters for either an
$R_y(\theta)$ gate or an $R_z(\alpha)$ gate to place into boxes labelled
$Y$ or $Z$ respectively.  We define
$\# \tau$ to be the total of the number of controlled-nots and
boxes, while $\mbox{dim} \tau$ denotes the number of boxes.
Label $S_\tau$ to be the subset of all $V \in U(N)$
that result from choosing particular parameters for a
$R_y(\theta)$ gate in each $Y$ box and an $R_z(\alpha)$ gate
in each $Z$ box.  We say that $\tau$ specializes stably to an analytic
subgroup $H \subset U(N)$ when $\tilde{S}_\tau \subset H$.

\begin{lemma}
\label{lem:measure0}
Suppose $\tau$ specializes stably to $H$ and
$\mbox{dim} \; \tau +1 < \mbox{dim} \; H$.  Then $\tilde{S}_\tau$
is a measure zero subset of $H$.
\end{lemma}

\begin{proof}
We appeal to Sard's theorem from differential topology
\cite[p.39]{GuilleminPollack:74}.
Consider the map $f: \mathbb{R}^{\mbox{dim} \; \tau + 1} \rightarrow
U(N)$ which carries a tuple $(\Phi, t_1, t_2, \cdots, t_{\mbox{dim} \; \tau
+ 1})$ to the $\mbox{e}^{i\Phi} V$ which is the phase $\mbox{e}^{i\Phi}$
multiplied by the specialization of $\tau$ corresponding to
$t_1,t_2,\cdots,t_{\mbox{dim}\; \tau}$.  This map is smooth.

By Sard's theorem \cite[p.39]{GuilleminPollack:74}, for all but
a measure zero subset of $h\in H$ one of the following two
cases hold:
\begin{itemize}
\item There is no choice of parameter $v$ with $f(v)=h$.
\item For each $v$ with $f(v)=h$, the
derivative linear map at the
parameter $v$ denoted $df_v : \mathbb{R}^{\mbox{dim}\; \tau + 1}
\rightarrow T_h(H)$ is onto.
\end{itemize}
The second possibility is absurd by the dimension hypothesis.  Thus
$f(\mathbb{R}^{\mbox{dim} \; \tau+1})$ is a measure zero subset of $H$.
\end{proof}

\begin{proposition}
Fix $n$, and let $\varsigma$ be a quantum circuit synthesis algorithm
inputting $a \in A(n)$ and outputting stably to $A(n)$.  Then
$\# \varsigma \geq 2^{n}-1 = N-1$.
\end{proposition}

\begin{proof}  Let $C$ be a countable set with
$\{ \tau(c) \; ; \; c \in C\}$ the set of topologies output
by $\varsigma$.  Now
$\mbox{dim} \; A(n) = N$.  Thus assume by way of contradiction
$\# \varsigma < N-1$.  Then we may write
\begin{equation}
A(n) = \cup_{c \in C} \tilde{S}_{\tau(c)}
\end{equation}
This is impossible by Lemma \ref{lem:measure0}.  Indeed,
a countable union of measure zero subsets
is still measure $0$ and hence can not cover $A(n)$.
\end{proof}

\begin{corollary}  Let $\{\varsigma(n)\}$ be a family of synthesis algorithms,
each of which accepts all inputs from  $A(n)$
and outputs stably to $A(n)$ per Definition
\ref{def:stablyasymptoticallyoptimal}.
If $\# \varsigma(n) \in O(2^n)$, then $\{ \varsigma(n)\}$
is stably asymptotically optimal.
\end{corollary}

\section{Conclusions and On-Going Work}
\label{sec:conclusions}

We realize quantum circuits for any diagonal
$U=\sum_{j=0}^{N-1} u_j \ket{j} \bra{j}$ consisting of at most
$2^{n+1}-3$ alternating controlled-not gates and $z$ axis Bloch sphere
rotations on individual qubits.
The construction uses
a new circuit block, the ${\tt XOR}$-controlled
rotation.  This $O(2^n)$ construction is
optimal in the following sense.  In the
worst-case and also the generic case, at least
$2^n-1$ one-qubit rotations are required to construct such a diagonal $U$.
Thus our constructive algorithm shows that the synthesis of quantum circuits
for diagonal computations is in fact $\Theta (2^n)$.  Note that special-case
computations such as tensors of one-qubit diagonal computations may require
fewer gates.

The circuits above have several common applications.  For example, they
are useful when constructing a circuit for a top-conditioned $V$
computation given a circuit diagram for $V$ correct up to relative phase.
They are also needed when applying projective measurements other than the
typical $\{ \bra{j} \; ; \; 0 \leq j \leq N-1\}$.
In our ongoing work, we will explore applications relating to the synthesis
of real quantum computations and also exotic quantum circuit synthesis
algorithms relying on $KAK$ metadecompositions of $U(2^n)$.

\appendix

\section{Synthesis via Controlled Rotations}
\label{sec:cr}

This appendix describes a synthesis algorithm
using the $\Lambda_k [R_z(\alpha)]$ circuit subblocks.
Recall our constructive proof of the upper bound on gate
counts of Theorem \ref{thm} used $\bigoplus_k [R_z(\alpha)]$
subblocks instead.
Several technical issues arising in our main algorithm
also arise here.  Thus, this appendix may serve as an introduction
of how to use the obstruction $\eta(-)$
of Definition \ref{def:chi&eta}
to form a recursive synthesis algorithm reducing $n$-qubit diagonals to
$(n-1)$-qubit diagonals.

\subsection*{Computation of $\eta(\; \Lambda_S[R_z(\alpha)] \; )$}

Recall from the Introduction $U=\sum_{j=0}^{N-1} u_j \ket{j} \bra{j}$
for $N=2^n$ a fixed $n$-qubit diagonal quantum computation.
Further recall that for $S \subset \{1,2,\cdots,n-1\}$,
by $\Lambda_S (V)$ for $V \in U(2^1)$ we mean
that instance of the $\# S$-conditioned computation
$\Lambda_{\#S}(V)$ which is conditioned on lines $\{ j \in S\}$
and acts on line $n$.

Every computation $\Lambda_S [R_z(\alpha)]$ is also diagonal.
We seek an explicit formula for
$\eta( \; \Lambda_S [R_z(\alpha)] \;)$.  With sufficient understanding
of how $\Lambda_S [R_z(\alpha)]$ affects $\eta(-)$, we will be able
to choose exact angles $\alpha$ so that preprending the conditioned
blocks to $U$ forces the composite to have $\eta = \vec{0}$.  Thus
the composite will be a tensor by corollary
\ref{cor:obstruction}, allowing for recursion.  The following
language is useful for expressing and computing
$\eta( \; \Lambda_S [R_z(\alpha)] \;)$.  It is slightly more convenient
to use the mathematical notation for vectors rather than {\tt bra-ket}.

\begin{definition}
For $1\leq j \leq N/2-1$, let $e_j$ denote the standard basis column
vectors
for $\mathbb{R}^{N/2-1}$, i.e., $e_j$ has
a single entry of $1$ in the $j^{\mbox{th}}$
row and all other entries $0$.
We further define the vectors $v_j=e_j-e_{j+1}$ if $1 \leq j \leq N/2-2$,
also setting $v_0=-e_1$ and $v_{N/2-1}=e_{N/2-1}$.
\end{definition}

Observe that the vectors $\{v_j \; ; 1 \leq j \leq N/2-1\}$
form a basis for $\mathbb{R}^{2^{n-1}-1}$.
We need one further convention to describe
$\eta( \; \Lambda_S [R_z(\alpha)] \;)$.

\begin{definition}
Let $1 \leq j \leq N/2-1$, with binary representation
$j=b_1 b_2 \cdots b_{n-1}$ for $b_1, b_2, \cdots, b_{n-1} \in \mathbb{F}_2$.
Let $S \subset \{1,2,\cdots, n-1\}$, $S \neq \emptyset$.
We say that $j$ is $S$-conditioned iff
$\prod_{j \in S} b_j=1$.  We label
$\mathcal{C}(S)=\{ j \; ; \; j \mbox{ is } S \mbox{ conditioned} \}$.
\end{definition}

\begin{figure}
\begin{center}
\begin{picture}(26,9)
\put(1,0){\usebox{\Rzgate}}
\put(1,2){\usebox{\hwire}}
\put(1,4){\usebox{\hwire}}
\put(1,6){\usebox{\hwire}}
\put(1,8){\usebox{\hwire}}
\put(0,9){$\ket{0}$}
\put(2,3){\circle*{0.4}}
\put(2,5){\circle*{0.4}}
\put(2,7){\circle*{0.4}}
\put(2,1.8){\line(0,1){5.2}}

\put(3.5,4.5){{\LARGE $=$}}

\put(7,3){\circle*{0.4}}
\put(7,5){\circle*{0.4}}
\put(7,7){\circle*{0.4}}
\put(7,9){\circle{1}}
\put(7,3){\line(0,1){6.5}}

\put(5,9){$\ket{0}$}
\put(6,0){\usebox{\hwire}}
\put(6,2){\usebox{\hwire}}
\put(6,4){\usebox{\hwire}}
\put(6,6){\usebox{\hwire}}
\put(6,8){\usebox{\hwire}}

\put(9,9){\circle*{0.4}}
\put(9,1.8){\line(0,1){7.2}}

\put(8,0){\usebox{\Rzgate}}
\put(8,2){\usebox{\hwire}}
\put(8,4){\usebox{\hwire}}
\put(8,6){\usebox{\hwire}}
\put(8,8){\usebox{\hwire}}

\put(11,3){\circle*{0.4}}
\put(11,5){\circle*{0.4}}
\put(11,7){\circle*{0.4}}
\put(11,9){\circle{1}}
\put(11,3){\line(0,1){6.5}}

\put(10,0){\usebox{\hwire}}
\put(10,2){\usebox{\hwire}}
\put(10,4){\usebox{\hwire}}
\put(10,6){\usebox{\hwire}}
\put(10,8){\usebox{\hwire}}

\put(16,3){\circle*{0.4}}
\put(16,5){\circle*{0.4}}
\put(16,7){\circle*{0.4}}
\put(16,9){\circle{1}}
\put(16,3){\line(0,1){6.5}}

\put(14,9){$\ket{0}$}
\put(15,0){\usebox{\hwire}}
\put(15,2){\usebox{\hwire}}
\put(15,4){\usebox{\hwire}}
\put(15,6){\usebox{\hwire}}
\put(15,8){\usebox{\hwire}}

\put(17.5,4.5){{\LARGE $=$}}

\put(21,1){\circle*{0.4}}
\put(21,3){\circle*{0.4}}
\put(21,9){\circle{1}}
\put(21,1){\line(0,1){8.5}}

\put(22,5){\circle*{0.4}}
\put(22,7){\circle*{0.4}}
\put(22,1){\circle{1}}
\put(22,7){\line(0,-1){6.5}}

\put(23,1){\circle*{0.4}}
\put(23,3){\circle*{0.4}}
\put(23,9){\circle{1}}
\put(23,1){\line(0,1){8.5}}

\put(24,5){\circle*{0.4}}
\put(24,7){\circle*{0.4}}
\put(24,1){\circle{1}}
\put(24,7){\line(0,-1){6.5}}

\put(19,9){$\ket{0}$}
\put(20,0){\usebox{\hwire}}
\put(20,2){\usebox{\hwire}}
\put(20,4){\usebox{\hwire}}
\put(20,6){\usebox{\hwire}}
\put(20,8){\usebox{\hwire}}

\put(22,0){\usebox{\hwire}}
\put(22,2){\usebox{\hwire}}
\put(22,4){\usebox{\hwire}}
\put(22,6){\usebox{\hwire}}
\put(22,8){\usebox{\hwire}}

\put(23,0){\usebox{\hwire}}
\put(23,2){\usebox{\hwire}}
\put(23,4){\usebox{\hwire}}
\put(23,6){\usebox{\hwire}}
\put(23,8){\usebox{\hwire}}

\end{picture}
\end{center}

\caption{
\label{fig:CRcircuit}
This diagram \cite[Lemma 7.11]{BarencoEtAl:95} illustrates how to realize a
$\Lambda_k [R_z(\alpha)]$ via a singly controlled rotation and
$k$-controlled-nots.
The latter may be synthesized
using $O(k)$ elementary gates, \emph{given the ancilla qubit shown
as the top line}.  Without the ancilla, a
$O(k^2)$ gates would be required per corollary 7.6 ibid.
The diagram at right recalls the next step of the decomposition.
}
\end{figure}
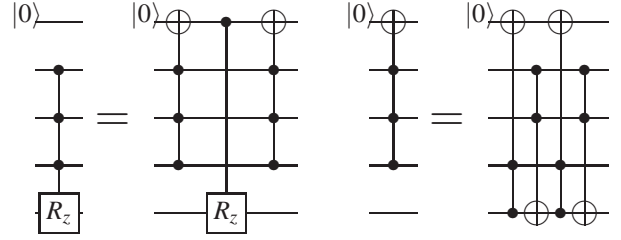

\begin{proposition}
\label{prop:LogChiCR}
Let $\mathcal{C}(S)$ denote the $S$-conditioned set
for some nonempty $S \subset \{1,\cdots,n-1\}$.  Then
\begin{equation}
\eta(\; \Lambda_S[R_z(\alpha)] \; ) =
\alpha \sum_{j \in \mathcal{C}(S)} v_j
\end{equation}
\end{proposition}

\begin{proof}
Label
$V=\Lambda_S[R_z(\alpha)] = \sum_{j=0}^{N-1} \lambda_j \ket{j} \bra{j}$.
We recall that
$\eta(V)$ is defined in terms
of $\chi_j(V)=
\lambda_{2j-2} \lambda_{2j-1}^{-1} \lambda_{2j}^{-1} \lambda_{2j+1}$.  Now
if $j \in \mathcal{C}(S)$, then
$\lambda_{2j}=\mbox{e}^{-i \alpha/2}$ and
$\lambda_{2j+1}=\mbox{e}^{i \alpha/2}$.
If the binary expression for $j$ is not $S$-conditioned, then
$\lambda_{2j}=\lambda_{2j+1}=1$.
Continuing in this manner, say the binary expression for
$j+1 \in \mathcal{C}(S)$.  Then
$\lambda_{2j+2}=\mbox{e}^{-i\alpha/2}$ and
$\lambda_{2j+3}=\mbox{e}^{i\alpha/2}$,
else $\lambda_{2j+2}=\lambda_{2j+3}=1$.  Thus letting
$\delta_{\mathcal{C}(S)}$ denote the indicator function of
$\mathcal{C}(S)$,
\begin{equation}
-i \log \chi_j (V) = \alpha \delta_{\mathcal{C}(S)} (j)
- \alpha \delta_{\mathcal{C}(S)}(j+1)
\end{equation}
This expression agrees componentwise with the result of the
proposition, given Definition \ref{def:chi&eta}.
\end{proof}

\begin{example}
\label{ex:computelogCR}
Consider $n=4$ qubits for the subset $S=\{1,3\}$ and $0 \leq \alpha < 2\pi$
 arbitrary.  Label $\phi = -\alpha/2$, so that
$R_z(\alpha) = \mbox{e}^{i \phi} \ket{0}\bra{0} +
\mbox{e}^{-i \phi} \ket{1}\bra{1}$.
Since $V=\Lambda_S[R_z(\alpha)]$ is diagonal, we describe
the quantum computation by specifying
multiples on each computational basis state.
\begin{center}
\small
\begin{tabular}{|l|c||l|c||l|c||l|c|}
\hline
state & mult & state & mult & state & mult & state & mult \\
\hline
\hline
$\ket{0000}$ & $1$ & $\ket{0100}$ & $1$ &
$\ket{1000}$ & $1$ & $\ket{1100}$ & $1$ \\
\hline
$\ket{0001}$ & $1$ & $\ket{0101}$ & $1$ &
$\ket{1001}$ & $1$ & $\ket{1101}$ & $1$ \\
\hline
$\ket{0010}$ & $1$ & $\ket{0110}$ & $1$ &
$\ket{1010}$ & $\mbox{e}^{i \phi}$ & $\ket{1110}$ & $\mbox{e}^{i\phi}$ \\
\hline
$\ket{0011}$ & $1$ & $\ket{0111}$ & $1$ &
$\ket{1011}$ & $\mbox{e}^{-i\phi}$ & $\ket{1111}$ & $\mbox{e}^{-i\phi}$ \\
\hline
\end{tabular}
\end{center}
Thus, $\chi_1(V)=1$,
$\chi_2(V)=1$, $\chi_3(V)=1$,
$\chi_4(V)=1$, $\chi_5(V)=\mbox{e}^{-2i\phi}$,
$\chi_6(V)=\mbox{e}^{2i\phi}$, and
$\chi_7(V)=\mbox{e}^{-2i\phi}$.  Thus we have directly computed that
$\eta( \; \Lambda_S [R_z(\alpha)] \;) =
-2 \phi i [0\ 0\ 0\ 0\ 1\ -1\ 1]^t$.

On the other hand,
$\mathcal{C}(\{1,3\})=\{101_{\mbox{b}},111_{\mbox{b}}\} = \{5,7\}$, where
the subscript denotes binary.
Thus
\begin{equation}
v_5+v_7=
(e_5-e_6)+e_7 = [0\ 0\ 0\ 0\ 1\ -1\ 1]^t
\end{equation}
Thus we computed the right-hand side of Proposition \ref{prop:LogChiCR}.
\end{example}

\subsection{$\Lambda_k[R_z(\alpha)]$-block synthesis algorithm}

Before the following definition, we note a happy accident.
There are $N/2-1$ nonempty subsets of the top lines
$\{1,\cdots,n-1\}$, and moreover $N/2-1$ characters
$\chi_j:A(n)\rightarrow U(1)$ which must be zeroed within the components
of the obstruction $\eta(-)$ to form a tensor.
Thus, the following matrix is square.

\begin{definition}
\label{def:LogCRMat}
The $(N/2-1) \times (N/2-1)$ real matrix ${\bf \eta}^\Lambda$ is
defined as follows.
Order nonempty subsets
$S_1$, $S_2$, \dots $S_{2^{n-1}-1}$ in dictionary order.  Then for
$1 \leq j \leq N/2-1$, the $j^{\mbox{th}}$ column of
${\bf \eta}^\Lambda$ is
$\eta (\; \Lambda_{S_j}[R_z(1 \; \mbox{radian})] \;)$.
\end{definition}

\begin{lemma}
\label{lem:CRcomp}
Let $\vec{\alpha}=[\alpha_1 \ \cdots \ \alpha_{N/2-1}]^t$.
Then for $S_1$, $S_2$, \dots $S_{N/2-1}$ the dictionary ordering
of nonempty subsets of $\{1,\cdots,n-1\}$,
\begin{equation}
\eta ( \; \Lambda_{S_1}[R_z(\alpha_1)] \;
\Lambda_{S_2} [R_z(\alpha_2)] \; \cdots \;
\Lambda_{S_{N/2-1}} [R_z(\alpha_{N/2-1})] \; ) =
{\bf \eta}^\Lambda \vec{\alpha}
\end{equation}
Here, the right hand side denotes matrix multiplication by the
column vector $\vec{\alpha}$.
\end{lemma}

\begin{sketch}  Recall that for any character
$\chi : A \rightarrow \mathbb{C}-\{0\}$, one has
$\log {\chi}(V W) =
\log {\chi}(V)+\log \vec{\chi}(W)$ and
$\log \chi (V^a) = a \log \chi (V)$ for $V,W \in A$, $a \in \mathbb{R}$.
Recall Definition \ref{def:chi&eta} and apply
these properties to the entries
$-i \log \vec{\chi}_j$ of the vector valued function $\eta(-)$.
\end{sketch}

We now state $\Lambda_k[R_z(\alpha)]$-block synthesis algorithm for a diagonal
unitary computations.  The proof of correctness
follows in the next subsection and includes a proof of the
subtle fact that the matrix ${\bf \eta}^\Lambda$ is inveritble.

\noindent
{\bf Controlled Rotation Synthesis Algorithm}
Let $U=\sum_{j=0}^{N-1} u_j \ket{j} \bra{j}$, for which
we wish to synthesize a circuit diagram in terms of
$\Lambda_k [R_z(\alpha)]$ blocks.
Label $S_1$, $S_2$, $S_3$ \dots $S_{2^{n-1}-1}$ the
nonempty subsets of the top $n-1$ lines $\{1,\cdots, n-1\}$ in dictionary
order.
\begin{enumerate}
\item{ Compute the obstruction $\vec{\psi}= \eta(U)$.}
\item{ Compute the inverse matrix
$({\bf \eta}^\Lambda)^{-1}$.}
\item{ Compute $\vec{\alpha}=
({\bf \eta}^\Lambda)^{-1} \vec{\psi}$, treating
$\vec{\psi}$ as a column vector.  Label
$\vec{\alpha}=[\alpha_1 \  \cdots \alpha_{N/2-1}]^t$.}
\item{ Compute the diagonal quantum computation
$\tilde{U}=
\Lambda_{S_1}[R_z(-\alpha_1)] \; \cdots \;
\Lambda_{S_{N/2-1}}[R_z(-\alpha_{N/2-1})] \; U$.  As is
verified below, $\tilde{U}$ is a tensor.}
\item{ Use the argument of prop. \ref{prop:tensor} to compute
$\tilde{U}=V \otimes W$ for $V \in A(n-1)$ and
$W=\mbox{e}^{i \Phi}R_z(\alpha_0)$ for some angle $\alpha_0$.}
\item{ Given prior computations, the following
expression holds:
\begin{equation}
U = \Lambda_{\emptyset}[R_z(\alpha_0)] \;
\Lambda_{S_1}[R_z(\alpha_1)] \;
\cdots \; \Lambda_{S_{N/2-1}}[R_z(\alpha_{N/2-1})] \; [V \otimes {\bf 1}]
\end{equation}
Here, ${\bf 1}$ denotes the trivial computation of $U(2^1)$.
Also, $\Lambda_\emptyset [R_z(\alpha_0)]$ means ${\bf 1} \otimes
R_z(\alpha_0)$ for ${\bf 1} \in U(2^{N/2})$.}
\item
Techniques from the literature are then used to decompose each
$\Lambda_{S_j}[R_z(\alpha_j)]$ into
elementary gates per Figure \ref{fig:CRcircuit}.
\item{
The algorithm terminates by recursively producing a circuit diagram for
$V \in A(n-1)$.}
\end{enumerate}

\vspace{-.2cm}

\begin{example}

In three qubits, consider the following diagonal computation.
\begin{equation}
\begin{array}{lcl}
U & = &
\mbox{e}^{6\pi i/6}\ket{0}\bra{0}+ \mbox{e}^{3 \pi i/6} \ket{1}\bra{1}+
\mbox{e}^{9\pi i/6}\ket{2}\bra{2}+ \mbox{e}^{8 \pi i/6}\ket{3}\bra{3} \\
& & + \mbox{e}^{5\pi i/6}\ket{4}\bra{4}+ \mbox{e}^{1 \pi i/6}\ket{5}\bra{5}+
\mbox{e}^{6\pi i/6}\ket{6}\bra{6}+ 1 \ket{7}\bra{7} \\
\end{array}
\end{equation}
Then one has
$\chi_1 (U) = \mbox{e}^{2 \pi i/6}$, $\chi_2(U)=\mbox{e}^{-3\pi i/6}$,
$\chi_3 (U) = \mbox{e}^{-2 \pi i/6}$ so that
$\vec{\psi}=\eta (U) = [ 2\pi/6 \ -3\pi/6 \ -2\pi/6]^t$.

We now must compute $\vec{\alpha}$ by computing the inverse matrix
$({\bf \eta}^\Lambda)^{-1}$.  For this matrix,
first compute the following.
\begin{equation}
{\bf \eta}^\Lambda
=
\left(
\begin{array}{ccc}
0 & 0 & 1 \\
1 & 0 & -1 \\
0 & 1 & 1 \\
\end{array}
\right)
\end{equation}
The following inverse matrix results, and it may be reused for multiple
specific diagonals $U$.
\begin{equation}
({\bf \eta}^\Lambda)^{-1}=
\left(
\begin{array}{ccc}
1 & 1 & 0 \\
-1 & 0 & 1 \\
1 & 0 & 0 \\
\end{array}
\right)
\end{equation}
So $\vec{\alpha} = ({\bf \eta})^{-1} \vec{\psi}
= [ -\pi/6 \ -4\pi/6 \; \ 2\pi/6 ]^t$.  Hence $\tilde{U}$
as defined below is a tensor.
\begin{equation}
\tilde{U} =
\Lambda_{\{1\}}[R_z(\pi/6)] \;
\Lambda_{\{1,2\}}[R_z(4\pi/6)] \; \Lambda_{\{2\}}[R_z(-\pi/6)] \; U
\end{equation}

In order to verify this directly, we compute the eight diagonal matrix
coefficients of each of $\Lambda_S[R_z(\alpha)]$.  To save space,
we write $\mbox{diag}(\lambda_0, \cdots, \lambda_7)$ for
$\lambda_0 \ket{0} \bra{0} + \cdots + \lambda_7 \ket{7} \bra{7}$.
\begin{equation}
\begin{array}{lcl}
\Lambda_{\{1\}}[R_z(\pi/6)] & = & \mbox{diag}(1,1,1,1,\mbox{e}^{-\pi i/12},
\\
& & \mbox{ \quad \quad}
\mbox{e}^{\pi i/12}, \mbox{e}^{-\pi i/12}, \mbox{e}^{\pi i/12}) \\
\Lambda_{\{1,2\}}[R_z(4 \pi/6)] & = & \mbox{diag}(1,1,1,1,1,1,
\mbox{e}^{-4 \pi i/12},
\mbox{e}^{4 \pi i/12}) \\
\Lambda_{\{2\}}[R_z(-2 \pi/6)] & = & \mbox{diag}(1,1,\mbox{e}^{2 \pi i/12},
\mbox{e}^{-2 \pi i/12}, \\
& & \mbox{ \quad \quad}
1,1,\mbox{e}^{2 \pi i/12},
\mbox{e}^{-2 \pi i/12}) \\
\end{array}
\end{equation}
Then multiplying, the expression demonstrates
$\tilde{U}=V\otimes W$.
\begin{equation}
\begin{array}{lcl}
\tilde{U} & = & \mbox{diag}(\mbox{e}^{12 \pi i/12},
\mbox{e}^{6 \pi i/12}, \mbox{e}^{20 \pi i/12}, \mbox{e}^{14 \pi i/12}, \\
& & \mbox{\quad \quad \quad}
\mbox{e}^{9 \pi i/12}, \mbox{e}^{3 \pi i/12}, \mbox{e}^{9 \pi i/12},
\mbox{e}^{3 \pi i/12}) \\
\end{array}
\end{equation}
Since $\tilde{U}$ is a tensor, we obtain the following decomposition of $U$.
\begin{equation}
\begin{array}{lcl}
U & = & \Lambda_{\{1\}}[R_z(-\pi/6)] \;
\Lambda_{\{1,2\}}[R_z(-4\pi/6)] \; \Lambda_ {\{2\}}[R_z(\pi/6)]
\;
\\ & &
[\mbox{diag}(1,\mbox{e}^{8 \pi i/12}, \mbox{e}^{-3 \pi i/12},
\mbox{e}^{-3 \pi i/12}) \otimes \mbox{diag}(\mbox{e}^{12 \pi i/6},
\mbox{e}^{6 \pi i/6})] \\
\end{array}
\end{equation}
The algorithm then recursively synthesizes the $2$-qubit diagonal
$V=1\ket{0}\bra{0}+\mbox{e}^{8 \pi i/12}\ket{1}\bra{1}+
\mbox{e}^{-3 \pi i/12}\ket{2}\bra{2} +
\mbox{e}^{-3 \pi i/12}) \ket{3}\bra{3}$.
\end{example}

\vspace{-.2cm}

\subsection*{Proof of correctness of $\Lambda_k[R_z(\alpha)]$-block synthesis}

We briefly verify that $\tilde{U}=V \otimes W$.  First use proposition
\ref{lem:CRcomp} for
\begin{equation}
\eta(\; \Lambda_{S_1}[R_z(-\alpha_1)] \; \cdots \;
\Lambda_{S_{N/2-1}}[R_z(-\alpha_{N/2-1})] \; ) = - \vec{\psi}
\end{equation}
Then the property $\eta(U_1 U_2) = \eta(U_1) + \eta(U_2)$ demands
\begin{equation}
\eta( \; \Lambda_{S_1}[R_z(-\alpha_1)] \; \cdots
\Lambda_{S_{N/2-1}}[R_z(-\alpha_{N/2-1})] \; U \; ) =
-\vec{\psi} + \vec{\psi} = \vec{0}
\end{equation}
So by the restatement of Proposition \ref{prop:tensor}, we have
$\tilde{U} = V \otimes W$.

The algorithm also uses the following proposition.

\begin{proposition}
\label{prop:CRMatInvert}
The matrix ${\bf \eta}^\Lambda$ per Definition \ref{def:LogCRMat}
is an invertible
$(2^{n-1}-1) \times (2^{n-1}-1)$ matrix.
\end{proposition}

\begin{sketch}
It suffices instead to consider the similar matrix corresponding
to a change of basis to $v_j$, $1 \leq j \leq N/2-1$ of
Definition \ref{def:vdef}.
Thus, if $B= [ v_1 \ v_2 \ \cdots \
v_{N/2-1} ]$ is the change of basis matrix, the matrix similar to
${\bf \eta}^\Lambda$ is
$M= B^{-1} {\bf \eta}^\Lambda B = (m_{jk})$.  Now
$m_{jk}=0$ if $j$ is not $S_k$-conditioned and
$m_{jk}=1$ if $j$ is $S_k$-conditioned.

$M$ is invertible since column operations reduce $M$ to a permutation
matrix.  Indeed, the last $e_{N/2-1}$ column may be used to clear all
other nonzero entries in the last row.  Then each of the columns corresponding
to $n-2$ element subsets retain a single nonzero entry, and the
corresponding rows may be cleared.
\end{sketch}

\end{document}